\documentclass[a4paper,11pt]{article}
\pdfoutput=1 % if your are submitting a pdflatex (i.e. if you have
\usepackage[noconfig]{refstyle}
\usepackage[dvipsnames]{xcolor}
\usepackage[hyperfootnotes=false, linktocpage=true, colorlinks, citecolor=blue, linkcolor=blue, urlcolor=Maroon]{hyperref}
\usepackage[page]{appendix}

\usepackage{float, array, xspace, amscd, amsthm, amsmath, amssymb,amsfonts}
\usepackage[margin=1in,footskip=0.25in]{geometry}
\usepackage{fancyhdr}
\usepackage{longtable}
\usepackage{tikz-cd}
\usepackage{comment}
\usepackage{slashed}
\usepackage{mathrsfs}
\usepackage{cite}
\usepackage[normalem]{ulem}

\newcommand\Tstrut{\rule{0pt}{2.6ex}}       % top strut
\newcommand\Bstrut{\rule[-1.1ex]{0pt}{0pt}} % bottom strut

\newtheorem{v-theorem}{Vanishing Theorem}

\newcommand{\tr}{{\rm tr}}

\newcommand{\be}{\begin{equation}}
\newcommand{\ee}{\end{equation}}
\def\bea#1\eea{\begin{align}#1\end{align}}

\newcommand{\beq}{\begin{equation}}
\newcommand{\eeq}{\end{equation}}
\newcommand{\ba}{\begin{array}}
\newcommand{\ea}{\end{array}}
\newcommand{\eea}{\end{eqnarray}}
\newcommand{\bean}{\begin{eqnarray*}}
\newcommand{\eean}{\end{eqnarray*}}
\newcommand{\eref}[1]{(\ref{#1})}

\newcommand{\vref}[1]{Vanishing~Theorem~\ref{#1}} %can also abbreviate to Theorem~\ref{#1}
\usepackage{color}

\def\fnote#1#2{\begingroup\def\thefootnote{#1}\footnote{#2}
     \addtocounter{footnote}{-1}\endgroup}

\newcommand{\mycaption}[1]{\caption{{\sf \small #1}}}

\begin{document}

\vspace{1cm}

\title{       {\Large \bf Generalized Vanishing Theorems for Yukawa Couplings in Heterotic Compactifications}}

\vspace{2cm}

\author{
Lara~B.~Anderson,${}^{1}$
James~Gray,${}^{1}$ \\
Magdalena Larfors,${}^{2}$
Matthew Magill,${}^{2}$ and
Robin Schneider ${}^{2}$
}
\date{}
\maketitle
\begin{center} {\small ${}^1${\it Department of Physics, 
Robeson Hall, Virginia Tech \\ Blacksburg, VA 24061, U.S.A.}\\[0.2cm]
       ${}^2${\it Department of Physics and Astronomy, Uppsala University,\\
       $~~~~~$ SE-751 20 Uppsala, Sweden.}}\\

\fnote{}{lara.anderson@vt.edu}
\fnote{}{jamesgray@vt.edu}
\fnote{}{magdalena.larfors@physics.uu.se}
\fnote{}{matthew.magill@physics.uu.se}
\fnote{}{robin.schneider@physics.uu.se}

\end{center}

\begin{abstract}
\noindent
Heterotic compactifications on Calabi-Yau threefolds frequently exhibit textures of vanishing Yukawa couplings in their low energy description. The vanishing of these couplings is often not enforced by any obvious symmetry and appears to be topological in nature. Recent results used differential geometric methods to explain the origin of some of this structure  \cite{Blesneag:2015pvz,Blesneag:2016yag}. A vanishing theorem was given which showed that the effect could be attributed, in part, to the embedding of the Calabi-Yau manifolds of interest inside higher dimensional ambient spaces, if the gauge bundles involved descended from vector bundles on those larger manifolds. In this paper, we utilize an algebro-geometric approach to provide an alternative derivation of some of these results, and are thus able to generalize them to a much wider arena than has been considered before. For example, we consider cases where the vector bundles of interest do not descend from bundles on the ambient space. In such a manner we are able to highlight the ubiquity with which textures of vanishing Yukawa couplings can be expected to arise in heterotic compactifications, with multiple different constraints arising from a plethora of different geometric features associated to the gauge bundle.

\end{abstract}

\thispagestyle{empty}
\setcounter{page}{0}
\newpage

\tableofcontents

\newpage
%%%%%%%%%%

\section{Introduction} \label{intro}

One of the more phenomenologically interesting features of compactifications of heterotic string theory is that they frequently exhibit textures of vanishing Yukawa couplings \cite{Strominger:1985ks,Greene:1986bm,Greene:1986jb,Candelas:1987se,Distler:1987gg,Distler:1987ee,Greene:1987xh,Candelas:1990pi,Distler:1995bc,Braun:2006me,Bouchard:2006dn,Anderson:2009ge,Anderson:2010tc,Buchbinder:2014sya,Blesneag:2015pvz,Blesneag:2016yag}. These are generically not associated to any known symmetry of the theory and appear to be topological in nature. Ideally, one would like to study the physical Yukawa couplings in discussing such an effect. Unfortunately, this would require the inclusion of normalization considerations arising from the matter field K\"ahler potential and this object is extremely difficult to compute. Some approximations to the K\"ahler potential do exist in the literature \cite{Gray:2003vw,Candelas:2016usb,McOrist:2016cfl,Blesneag:2018ygh,Candelas:2018lib}, and its computation  is the ultimate goals of much of the work on numerical approaches to Ricci-flat metrics and gauge bundle connections on Calabi-Yau manifolds \cite{Donaldson,Headrick:2005ch,Douglas:2006hz,Douglas:2006rr,Braun:2007sn,Braun:2008jp,Headrick:2009jz,Anderson:2010ke,Anderson:2011ed,Ashmore:2019wzb,Cui:2019uhy,Anderson:2020hux,Douglas:2020hpv,Jejjala:2020wcc}. Nevertheless, it is fair to say that there is still very little  known about the physically normalized Yukawa couplings.
Despite this, the textures of vanishing Yukawa couplings mentioned above can be observed by considering the Yukawa coupling terms that appear in the four dimensional ${\cal N}=1$ superpotential. The goal of this paper is to shed more light on the origins of this observed structure.

Consider a compactification of heterotic string or M-theory which corresponds to a vector bundle $V_X$ over a smooth Calabi-Yau threefold $X$. Matter fields in the heterotic theory are in one to one correspondence with classes in the first cohomology groups valued in certain associated bundles to $V_X$. Such cohomology group elements can be represented in terms of bundle valued 1-forms which we will denote as $\nu_i$ where $i$ runs over the matter fields of interest. The superpotential Yukawa coupling between three such fields is then well known to be given by the following expression.
\begin{eqnarray} \label{yuk1}
W_{\textnormal{Yukawa}} = \int_X \Omega \wedge \tr \left( \nu_1\wedge \nu_2 \wedge \nu_3 \right)
\end{eqnarray}
In this expression, $\Omega$ is the nowhere vanishing holomorphic 3-form and the trace denotes the formation of a gauge singlet from the gauge indices on the 1-forms. Due to its quasi-topological nature, the integral in (\ref{yuk1}) can be evaluated using any 1-forms which are elements of the cohomology class associated to the matter fields of interest, with the same result being obtained regardless of choice of element. In particular, harmonic bundle valued 1-forms need not be used. 

The Yukawa coupling (\ref{yuk1}) can be equally well expressed in an algebraic manner. The quantity $\tr \left( \nu_1\wedge \nu_2 \wedge \nu_3 \right)$ appearing in that formula is an element of $H^3({\cal O}_X)$ and, given that this cohomology is one dimensional, is therefore proportional to $\overline{\Omega}$. The constant of proportionality that appears is the superpotential Yukawa coupling, up to some overall normalization which is the same for all such interactions. Although all of the methods we will describe in this paper generalize to other cases, we will restrict our considerations to bundles $V_X$ with structure group $SU(3)$. In such a case, then, the Yukawa coupling can be simply presented as a mapping. 
\begin{eqnarray} \label{mrmappy}
H^1( V_X) \otimes H^1(V_X) \otimes H^1(V_X) \to H^3(\wedge^3 V_X) \cong H^3({\cal O}_X) \cong \mathbb{C}
\end{eqnarray}
That is, the Yukawa coupling is a trilinear mapping (essentially a cup product composed with an anti-symmetrization) which takes three copies of the cohomologies associated with the families to a complex number. What has been observed in the literature is that the expressions (\ref{yuk1}) and (\ref{mrmappy}) frequently lead to vanishing couplings upon direct computation, even in cases where no obvious symmetry is present that would forbid the associated term in the superpotential (as just some examples of this see \cite{Braun:2006me,Bouchard:2006dn,Blesneag:2015pvz,Blesneag:2016yag,Gray:2019tzn}). 

There are two possibilities given the existence, and relative ubiquity \cite{Gray:2019tzn}, of such vanishing Yukawa couplings. The first is that this truly is a topological effect and it is indeed true that no associated symmetry is present. If this is the case then these textures will survive to low energy scales, not as vanishing couplings but rather as hierarchies, with those couplings that were present perturbatively at the compactification scale being large relative to those that were not. This effect could be important in explaining features such as the hierarchy in mass of the generations of the standard model, amongst others. Indeed, many heterotic compactifications can be ruled out by such vanishing Yukawa couplings making it impossible to obtain a sufficiently heavy top quark.

The second possibility is that these vanishings are secretly due to some symmetry, and that, in some cases, that symmetry simply hasn't been discovered yet. Indeed, in the relatively common case where a Yukawa coupling vanishes everywhere in complex structure moduli space, this possibility fits the spirit, if not the letter, of some recent swampland discussions \cite{Palti:2020qlc}. A coupling between families vanishing everywhere in moduli space would involve the vanishing of an infinite number of interaction terms in some sense. Specifically, couplings between those three representations of the gauge group and all possible powers of all of the complex structure moduli must vanish. Given this, reasoning such as that in \cite{Palti:2020qlc}, might lead one to suspect that such vanishings should not occur without some underlying cause related to symmetries. If this second possibility is indeed realized, then clearly the identification of these mysterious symmetries is a priority in the theory of heterotic compactifications. 

\vspace{0.1cm}

Given the above general discussion, it is important to learn as much as possible about the detailed structure of these vanishing Yukawa couplings and how they arise topologically. Recently, a differential approach, based on (\ref{yuk1}), was used to explain the topological origin of some vanishing Yukawa couplings \cite{Blesneag:2015pvz,Blesneag:2016yag}. The authors of that work considered cases where structures on the Calabi-Yau manifold descended nicely from related objects on a larger ambient space in which the threefold was embedded. The presence of this structure is what led to the vanishing of the Yukawa couplings. The current paper explains much of the structure of those vanishing theorems in an algebraic setting using (\ref{mrmappy}). In doing so, we are also able to greatly generalize that work in several directions, removing restrictions that are unavoidable in a differential approach. For example, we can deal with cases where the bundle does not descend from an equivalent structure on an ambient space.

As was mentioned above, to illustrate the underlying structure that we wish to emphasize, while minimizing technical complexity, we restrict our considerations in this paper to $SU(3)$ bundles $V_X$. Analogous results can certainly be derived for other structure groups. We then present a vanishing theorem on Yukawa couplings between the families of $V_X$ that is rather generally phrased.

Consider a case where $V_X$ appears at the right hand end of an exact sequence of sheaves (not necessarily vector bundles) of any length 
\begin{eqnarray} \label{introseq}
\ldots \to {\cal F}_2 \to {\cal F}_1 \to {\cal F}_0 \to V_X \to 0
\end{eqnarray}
Then all of the elements of $H^1(V_X)$ associated with the families of the associated four dimensional effective theory descend from an element of $H^i({\cal F}_{i-1})$ for some $i$. \emph{We call a field type $i$ when it is associated to an $i$'th cohomology group in this manner}. Then we can make the following statements.
\begin{itemize}
\item If $H^3(\wedge^3 {\cal F}_0)=0$  then all Yukawa couplings between three type 1 families vanish.
\item If in addition $H^4({\cal F}_1 \otimes \wedge^2 {\cal F}_0)=0$ then all Yukawa couplings between one type 2 and two type 1 families also vanish.
\end{itemize}

A theorem such as this, which depends upon a set of cohomologies vanishing, may not seem particularly generic, but in practice it is frequently extremely restrictive for the types of examples that are considered in the physics literature. One reason for this is that zeros in cohomology {\it do} frequently appear in examples. This is partially because, for an $SU(3)$ bundle $V_X$, $h^*(\wedge^3 V_X)=h^*({\cal O}) $. Therefore, sequences derived from (\ref{introseq}) containing $\wedge^3 V_X$ place strong constraints on the possible cohomology of combinations of the sheaves ${\cal F}_i$, simply due to the sparsity of the cohomology of the trivial bundle. Perhaps more importantly, however, {\it the vast majority of bundles $V_X$ considered in the physics literature admit a plethora of different sequences of the form (\ref{introseq}), any one of which has the potential to constrain the Yukawa couplings}.

To give an example of this, consider the case where the sequence (\ref{introseq}) is taken to be the Koszul sequence.
\begin{eqnarray} \label{introkos}
0 \to \wedge^k {\cal N}^{\vee} \otimes V\to \dots \to {\cal N}^{\vee} \otimes V \to V \to V_X \to 0
\end{eqnarray}
Here the Calabi-Yau $X$ has been taken to be described as a complete intersection of codimension $k$ in some ambient space ${\cal A}$ with normal bundle ${\cal N}$. The bundle $V_X$ has been taken to be the restriction of some sheaf\footnote{Subject to certain constraints that will be described in the main text.} $V$. First, the Calabi-Yau could be embedded in many different ambient spaces, leading to many different possible descriptions of the type (\ref{introkos}) \cite{Blesneag:2015pvz,Blesneag:2016yag}. Second, even for a given ambient space, the bundle $V_X$ could be obtained by restricting many different sheaves $V$. In principle all of these possibilities could lead to different constraints on the Yukawa couplings, and indeed in Section \ref{secegs} we will give concrete examples where precisely this phenomenon occurs.

It should be emphasized that Koszul sequences are just one choice for (\ref{introseq}), and many other possibilities probably exist. Even just given all of the readily available resolutions of the form (\ref{introseq}) described in the previous paragraph, it is perhaps not surprising that when Yukawa couplings are computed directly in examples in the literature, vanishings are often discovered.

\vspace{0.1cm}

The rest of this paper is structured as follows. In Section \ref{arg1} we review the differential geometric arguments of \cite{Blesneag:2015pvz,Blesneag:2016yag} which explain the structure of vanishing Yukawa couplings observed in some compactifications associated with a bundle $V_X$ descending from a bundle on some ambient space. In Section \ref{arg2} we provide a more general, algebraic, analysis of where vanishing Yukawa couplings can arise from, linking back to the case considered in Section \ref{arg1} to illustrate our statements. Section \ref{secegs} contains a number of different explicit examples to which the results of Section \ref{arg2} are applied. These examples illustrate an array of different compactification structures and show how different types of sequences of the form (\ref{introseq}) can interact in a single example to lead to highly constrained Yukawa couplings. In Section \ref{comparisonsec} we briefly discuss the relationship of our results to some other sources of vanishing Yukawa couplings in heterotic compactifications. Finally, in Section \ref{concsec} we briefly conclude our discussion.

\section{Vanishing Theorems from Differential Geometry} \label{arg1}

In this section we will review the basic argument of \cite{Blesneag:2015pvz,Blesneag:2016yag}. We will focus here on the situation where the Calabi-Yau threefold $X$ is described as an anti-canonical hypersurface, with defining relation $p$, in some ambient space ${\cal A}$. A more general analysis was performed for the case of higher codimension in \cite{Blesneag:2016yag}. Given the similarity of the derivations, we will only give details in codimension one and will simply content ourselves with a statement of the result in the more general case.

The authors of \cite{Blesneag:2015pvz,Blesneag:2016yag} consider the case where the bundle of interest $V_X$ is the restriction of a bundle $V$ on ${\cal A}$. In fact, those authors restrict their attention to sums of line bundles\footnote{A motivation to study such sums of line bundles is that they are often used in heterotic line bundle standard model constructions. See, for example \cite{Anderson:2011ns,Anderson:2012yf,Anderson:2013xka,He:2013ofa,Buchbinder:2013dna,Constantin:2015bea,Braun:2017feb,Otsuka:2018oyf,Constantin:2018xkj,Larfors:2020ugo,Otsuka:2020nsk,Deen:2020dlf,Larfors:2020weh}.}, but their proof holds for arbitrary structure group. We will refer to a $V_X$ associated to an ambient bundle in this way as being {\it ``ambiently defined''}. They then take the bundle valued 1-forms $\nu_i$ on $X$, corresponding to the matter fields of interest, to be the restriction of bundle valued 1-forms $\hat{\nu}_i$ on ${\cal A}$. Since, in the situation at hand, $\Omega$ can also be obtained as the restriction of ambient space form $\hat{\Omega}$ one can manipulate the expression for the Yukawa coupling (\ref{yuk1}) as follows
\begin{eqnarray} 
W_{\textnormal{Yukawa}} &=& \int_X \Omega \wedge \tr \left( \nu_1\wedge \nu_2 \wedge \nu_3 \right) \\
&=& \int_{\cal A} \hat{\Omega} \wedge \tr( \hat{\nu_1}\wedge\hat{\nu_2}\wedge \hat{\nu_3}) \wedge \delta^2(p) dp \wedge d\overline{p} \; .
\end{eqnarray}
Here the delta function with support over $p$ has been introduced so that we can replace the integral over $X$ in (\ref{yuk1}) with one over the ambient space. Applying the Poincar\'e-Lelong type identity
\begin{eqnarray}
\delta^2(p) d\overline{p} = \frac{1}{\pi} \overline{\partial}\left( \frac{1}{p} \right) \;,
\end{eqnarray}
we then arrive at the following expression
\begin{eqnarray} \label{yuk3}
W_{\textnormal{Yukawa}} = \int_{\cal A} \hat{\Omega} \wedge\tr( \hat{\nu_1}\wedge\hat{\nu_2}\wedge \hat{\nu_3}) \wedge dp \wedge \frac{1}{\pi} \overline{\partial}\left( \frac{1}{p} \right) \; .\end{eqnarray}
At this stage an integration by parts can be performed. One might worry about potential boundary terms in such a computation, but they have been shown to vanish in the detailed analysis of \cite{Blesneag:2016yag,Blesneag:2015pvz}, and as such we will not mention them further here. Given this, (\ref{yuk3}) becomes
\begin{eqnarray} \label{yuk4}
W_{\textnormal{Yukawa}} = - \frac{1}{\pi} \int_{\cal A} \hat{\Omega} \wedge dp  \wedge \frac{1}{p}  \overline{\partial} \,\tr( \hat{\nu_1}\wedge\hat{\nu_2}\wedge \hat{\nu_3})  \; .
\end{eqnarray}

\vspace{0.1cm}

To simplify this expression further, we need a simple form for the $\overline{\partial} \hat{\nu}_i$. Consider the Koszul sequence for the bundle of interest
\begin{eqnarray} \label{koszul1}
 0 \to {\cal N}^{\vee} \otimes V \stackrel{p}{\longrightarrow} V \stackrel{r}{\longrightarrow} V_X \to 0 \; .
\end{eqnarray}
Here, ${\cal N}$ is the normal bundle to $X$ in ${\cal A}$ so that $p \in H^0({\cal A},{\cal N})$. The maps $p$ and $r$ in (\ref{koszul1}) are given by the defining relation and restriction respectively. The form $\overline{\partial} \hat{\nu}_i \in \Omega^2(V)$ need not be zero, i.e. $\hat{\nu}_i$ need not be an element of $H^1({\cal A}, V)$. In fact, an examination of the long exact sequence associated to (\ref{koszul1}) reveals that the forms $\hat{\nu}_i$ can be separated into the two following types. Either $\nu_i$ is associated with an element of $H^1({\cal A}, V)$, which the authors refer to as a type 1 field, or $\nu_i$ is associated with an element of $H^2({\cal A},{\cal N}^{\vee} \otimes V)$, a type 2 field. In the former case $\overline{\partial} \hat{v}_i=0$. Therefore, we just need an expression for the form of $\overline{\partial} \hat{v}_i$ in the type 2 case.

Consider $r(\overline{\partial} \hat{\nu}_i)$ in the type 2 case\footnote{We are abusing notation somewhat here, in cases where it should not cause confusion, by using the same symbols for maps between bundles and the maps that they induce on bundle valued forms.}. We have that $r(\overline{\partial}\hat{\nu}_i) = \overline{\partial} (r(\hat{\nu}_i)) = \overline{\partial} \nu_i=0$. Thus $\overline{\partial}\hat{\nu}_i$ is in the kernel of $r$. Exactness of (\ref{koszul1}) then tells us that this form is also in the image of $p$. Thus we have that $\overline{\partial} \hat{\nu}_i = p \hat{\omega}_i$ for some $\hat{\omega}_i\in \Omega^2({\cal N}^{\vee}\otimes V)$. In fact $\hat{\omega}_i$ is closed, as can be seen from $p \overline{\partial} \hat{\omega}_i = \overline{\partial} (p \hat{\omega}) = \overline{\partial}^2 \hat{\nu}_i=0$ and the fact that $p$ is injective. (The above steps are exactly the derivation of the co-boundary map in the long exact sequence in cohomology associated to (\ref{koszul1}).) Finally then, we obtain a simple expression for $\overline{\partial} \hat{v}_i$ in the type 2 case. We have
\begin{eqnarray}
\overline{\partial} \hat{\nu}_i = p \hat{\omega}_i \;,
\end{eqnarray}
where $\hat{\omega}\in H^2({\cal A},{\cal N}^{\vee} \otimes V)$.

\vspace{0.1cm}

The results of the proceeding paragraph can then be used to further analyse (\ref{yuk4}). In particular, in the case where all of the fields in the coupling are of type 1, we have that $\overline{\partial} \hat{v}_i=0 \;\; \forall i$. In such a case, the $\overline{\partial} \,\tr( \hat{\nu_1}\wedge\hat{\nu_1}\wedge \hat{\nu_1})$ factor in (\ref{yuk4}) clearly gives zero, leading to a vanishing Yukawa coupling. If at least one of the fields is of type 2, however, the situation is different. Take the case where $\nu_1$ is of type 2 for example. Then we have,
\begin{eqnarray}
W_{\textnormal{Yukawa}} &=&- \frac{1}{\pi} \int_{\cal A} \hat{\Omega} \wedge dp  \wedge \frac{1}{p}  \,\tr( \overline{\partial}(\hat{\nu_1})\wedge\hat{\nu_2}\wedge \hat{\nu_3}+\ldots) \\
&=&- \frac{1}{\pi} \int_{\cal A} \hat{\Omega} \wedge dp  \wedge  \,\tr(  \hat{\omega}_1\wedge\hat{\nu_2}\wedge \hat{\nu_3}+\ldots) \;,
\end{eqnarray}
which is a coupling that is not necessarily vanishing. In the codimension one case, then, the authors of \cite{Blesneag:2015pvz}, find that all Yukawa couplings between three type 1 fields vanish. In \cite{Blesneag:2016yag} the authors generalized this result, using a similar analysis, to arbitrary codimension. Let us recapitulate these results here, but in a slightly more general form:

\begin{v-theorem} \label{v-theorem1}
Given a bundle $V_X$, on a complete intersection Calabi-Yau $X$, that descends from a bundle $V$ on the ambient space ${\cal A}$, we can define a notion of `type' for elements of $H^1(V_X)$. Namely, we say that such a cohomology element has type $\tau=i$ if it descends from an element of $H^{i}({\cal A},\wedge^{i-1}{\cal N}^{\vee} \otimes V_0)$. Then,
\begin{eqnarray} \label{th1}
\tau_1 +\tau_2 + \tau_3 < \textnormal{dim}{\cal A} \Rightarrow \lambda_{123} =0 \;,
\end{eqnarray}
where $\lambda_{ijk}$ is the Yukawa coupling between the indicated fields.
\end{v-theorem}

It should be emphasized that the work of \cite{Blesneag:2015pvz,Blesneag:2016yag} provides us with much more than this vanishing theorem. For example, their formalism enables the holomorphic Yukawa coupling to be computed as a function of the moduli of the problem. Nevertheless, in this paper we will focus on generalizing  \vref{v-theorem1} to a wider range of situations, in addition to comparing the textures caused by these types of vanishings to those arising from different sources of zeros. To achieve this, we need to rephrase the derivation of the vanishing result in algebraic terms.

\section{Vanishing Theorems from Algebraic Geometry} \label{arg2}

The vanishings seen in the previous section can be rephrased in purely algebraic terms. Such a reframing of this structure exhibited by the Yukawa couplings has both benefits and drawbacks. We will be able to generalize the result \vref{v-theorem1} to a very wide range of situations. In particular, we will not be tied to the structure of Koszul sequences and ambiently defined bundles, which are integral to the approach of the preceding section. A disadvantage of the algebraic approach will be that it will not be obvious in many situations how to practically obtain the moduli dependence of couplings being studied.

The constructions we will use in this section are standard within homological algebra. In particular, these techniques are at the heart of how the cup product underlying the algebraic definition of the Yukawa coupling (\ref{mrmappy}) is defined. That such technology is necessary is hardly surprising given that our goal is essentially to understand sub-structure of such cup products in various situations. However, this technology may be less familiar to some readers with a physics background. As such we have tried to be somewhat pedagogical in what follows, and the reader who does not wish to see the details of the derivations can simply skip to the result which is presented in Vanishing Theorem \ref{v-theorem4}. Readers interested in a more comprehensive review of homological algebra are referred to \cite{weibel}.

\subsection{Sequence chasing via homological algebra} \label{seqchase}

We will start our algebraic rephrasing of Yukawa coupling vanishing theorems by reproducing a very well known result using techniques of homological algebra. This will allow us to define our notation, introduce some relevant mathematical tools, and describe the objects that will replace the ambient space forms $\hat{\nu}$, which appeared in the discussion of Section \ref{arg1}, in this algebraic setting. 

\vspace{0.1cm}

Consider a situation where $V_X$ admits a short exact left resolution, that is where it appears as the right most object of a short exact sequence of sheaves
\begin{eqnarray} \label{seq1}
0 \to {\cal F}_1 \to {\cal F}_0 \to V_X \to 0 \; .
\end{eqnarray}
For example, if we wanted to make contact to the case considered in Section \ref{arg1}, we could use the Koszul sequence which arises if a bundle is ambiently defined in codimension one
\begin{eqnarray} \label{kos1}
0\to {\cal N}^{\vee} \otimes V \to V \to V_X \to 0 \; .
\end{eqnarray}
Note that although $V_X$ is a bundle on the Calabi-Yau threefold $X$, it is, in general, a sheaf on the ambient space ${\cal A}$ where this sequence is defined. As such, it will be essential in what follows that we consider the case where the sheaves appearing in our sequence (\ref{seq1}) are not necessarily locally free.

 Given any exact sequence of sheaves such as (\ref{seq1}) we can form a diagram as follows.
\begin{eqnarray} \label{quasi1}
\begin{array}{ccccccc} 0&\to &0&\to& V_X &\to& 0\\  &&\uparrow&&\uparrow&\\ 0 & \to & {\cal F}_1 & \to & {\cal F}_0& \to & 0\end{array}
\end{eqnarray}
Here we interpret the vertical maps between the two rows as forming a map between complexes $[ {\cal F}_1 \to {\cal F}_0] \to [V_X]$\footnote{We will use a notation $[X\to Y \to Z]$ for sequences where $0$'s are understood to be attached to the left and right and only non-zero entries are indicated.}. This map is a quasi-isomorphism of the complexes. That is, just by using the exactness of (\ref{seq1}), one can see that the homology of the top and bottom sequences in (\ref{quasi1}) are isomorphic.

A right resolution of a sheaf ${\cal F}$ is some exact sequence that has the sheaf at the left hand end
\begin{eqnarray}
 0\to {\cal F} \to  {\cal V}_0 \to {\cal V}_1 \to  {\cal V}_2 \to \ldots \; .
\end{eqnarray}
This is commonly denoted simply by $[{\cal F}\to {\cal V}_{\bullet} ]$. The two types of right resolution we will apply are injective resolutions and, what is in our context a special case of these, Godement resolutions. Injective resolutions have a few key properties that we will use in what follows. Firstly, every sheaf admits an injective resolution. Secondly, an injective resolution can be used to define the cohomology of a sheaf. Writing $[{\cal F} \to {\cal I}_{\bullet}({\cal F}) ]$ for the injective resolution, the homology of the sequence of global sections $[\Gamma({\cal I}_{\bullet}({\cal F}))]$ is the definition of the cohomology $H^{\bullet}({\cal F})$ of the sheaf ${\cal F}$. The last property of injective resolutions that we will use involves the existence of maps into them - we will discuss this as and when we need it.

If we have a map of complexes as in (\ref{quasi1}) we can apply various types of resolutions to all of the sheaves that appear. This results in a three-dimensional grid of objects which can be viewed as a map of bi-complexes. We can write
\begin{eqnarray} \label{quasi2}
\begin{array}{ccccccc} 0&\to &0&\to& G_{\bullet}(V_X )&\to& 0\\  &&\uparrow&&\uparrow&\\ 0 & \to &G_{\bullet}( {\cal F}_1) & \to & G_{\bullet}({\cal F}_0)& \to & 0\end{array}
\end{eqnarray}
where, for later use, we have used Godement resolutions $G$ - a special type of injective resolution whose features will be useful later on. This three dimensional grid of sequences can be viewed as a map between two two dimensional grids, or `bi-complexes', in analogy to the interpretation of (\ref{quasi1}). This can be denoted as $[G_{\bullet}({\cal F}_1) \to G_{\bullet}({\cal F}_0)] \to [G_{\bullet}(V_X)]$ and the  resulting map is a quasi-isomorphism at a constant level in the resolutions, that is, at any given subscript on the $G$'s. This is because the Godement resolution preserves exactness of short exact sequences, and so applying $G_i$ to (\ref{seq1}) gives a short exact sequence that can be built into such a map of complexes, just as we did in going from (\ref{seq1}) to (\ref{quasi1}).

The last piece of technical toolkit we will need in this subsection is the notion of taking a total complex of a bi-complex such as $[G_{\bullet}({\cal F}_{\bullet})]$. The total complex is a single complex whose entries are $\textnormal{Tot}([G_{\bullet}({\cal F}_{\bullet})])_q= \bigoplus_{i-j=q} G_{i}({\cal F}_j)$,\footnote{The minus sign appearing in this definition is a consequence of the resolutions $G_{\bullet}$ and ${\cal F}_{\bullet}$ being of different types (right and left respectively). If both of the sequences were of the same type, then the total complex would be defined with a plus sign.} and whose maps are built out of the maps in the original bi-complex in a specific manner. A crucial result in homological algebra states that if a map between bi-complexes is a quasi-isomorphism in its rows, such as was the case for the one in the preceding paragraph, then the induced morphism between the associated total complexes is a quasi-isomorphism of complexes \cite{weibel}. If we use this on the morphism $[G_{\bullet}({\cal F}_1) \to G_{\bullet}({\cal F}_0)] \to [G_{\bullet}(V_X)]$, then we obtain a quasi-isomorphism between the sequence
\begin{eqnarray} \label{one}
0 \to G_{0}({\cal F}_1) \to G_{1}({\cal F}_1) \oplus G_{0}({\cal F}_0) \to G_{2}({\cal F}_1) \oplus G_{1}({\cal F}_0) \to G_{3}({\cal F}_1) \oplus G_{2}({\cal F}_0) \to \ldots
\end{eqnarray}
and
\begin{eqnarray} \label{two}
0\to 0\to G_0 (V_X) \to G_1 (V_X) \to G_2 (V_X) \to \ldots
\end{eqnarray}
If we compare the homology of the global sections of (\ref{one}) and (\ref{two}) we obtain expressions for the cohomology $H^{\bullet}(V_X)$ in terms of structure appearing in (\ref{one}). 

To see how this works, consider the case of the Koszul sequence (\ref{kos1}) where we have ${\cal F}_0=V$ and ${\cal F}_1={\cal N}^{\vee} \otimes V$. In this case, comparing the cohomology of global sections of (\ref{one}) and (\ref{two}) gives exactly the same answer as sequence chasing the long exact sequence in cohomology associated to (\ref{kos1}). Indeed, this is essentially one way that that long exact sequence can be derived. So, for example, comparing the homology of global sections at the third entry (corresponding to $q=1$) in (\ref{one}) and (\ref{two}) in this case one quickly finds that 
\begin{eqnarray} \label{coh1}
H^1(V_X) = \ker \left\{ H^2({\cal N}^{\vee} \otimes V) \to H^2(V)\right\} \oplus \textnormal{coker}\left\{ H^1({\cal N}^{\vee} \otimes V) \to H^1(V) \right\} \; ,
\end{eqnarray}
which is exactly the result we were utilizing earlier in Section \ref{arg1}.

\subsubsection{Type and an ambiguity}

We can now discuss in more detail what is meant by type in this language. The fact, mentioned above, that Godement resolutions preserve exactness of short exact sequences means that the vertical map $[G_{\bullet}({\cal F}_0) ]\to [G_{\bullet}(V_X)]$ on the right of (\ref{quasi2}) is surjective. Therefore any section of $G_1(V_X)$, including those which correspond to elements of $H^1(V_X)$, has some pre-image in the sections of $G_1({\cal F}_0)$ (or $G_1(V)$ if we are utilizing the Koszul resolution (\ref{kos1})). If the element of $H^1(V_X)$ is type 1 then this pre-image will be associated to an element of cohomology. If it is of type 2 then the preimage will not be associated to an element of cohomology. Roughly speaking, the notion of the lifted form $\hat{\nu}$ in Section \ref{arg1} is replaced, in algebraic language, by sections of $G_1(V)$. Note that this form of `lifting' works for any resolution (\ref{seq1}), not just the specific case (\ref{kos1}).

It will be useful for the following sections of the paper to be a little more specific about the form of sections of Godement sheaves that correspond to elements in the cohomologies appearing in (\ref{coh1}). Consider $(w,v) \in \Gamma(G_2({\cal F}_1)) \oplus \Gamma(G_1({\cal F}_0))$, a section of the third entry in (\ref{one}), where $w\neq0$. Under the map induced from the total sequence (\ref{one}), this is mapped to $(dw,dv-f(w)) \in \Gamma(G_3({\cal F}_1)) \oplus \Gamma(G_2({\cal F}_0))$ (here $d$ is the map associated to the Godement sequence and $f$ that associated to the resolution (\ref{seq1})). Therefore, for $(w,v)$ to be in the kernel of the map, we require that $w$ is closed and that $ f(w)=dv $. This corresponds to an element of the kernel in (\ref{coh1}). The remaining structure necessary to correctly identify elements in the same cohomology class comes from the quotient of this kernel in the total sequence by the appropriate image to form the homology. For later use it will be important to note that this kind of type 2 element is therefore represented by a pair of sections of Godement sheaves. Indeed, the $v\in \Gamma(G_1({\cal F}_0))$ obeying $dv = f(w)$ is precisely the non-closed `lift' associated to the element of $H^1(V_X)$ being described, as was discussed in the previous paragraph.

By contrast if $w=0$ then $(0,v)$ is mapped to $(0,d v)$. This describes an element of $H^1(V)$ in the cokernel appearing in (\ref{coh1}) with the quotienting associated to taking homology leading to both the cokernel condition and the remaining structure required to define the classes in the cohomology appearing in (\ref{coh1}). Thus each element in a basis of the cokernel in (\ref{coh1}) can be written in terms of just a single section of a single Godement sheaf, $G_1({\cal F}_0)$, with no further structure.

An important consequence of the above discussion is that there is no such thing as a ``pure" type 2 field. We see this from the above description in that any closed section of $G_1({\cal F}_0)$ can be added to the $v$ associated to a type 2 field and the condition $f(w)=dv $ will still be satisfied. In other words, there is an ambiguity in defining the type 2 field in this language which corresponds exactly to adding an arbitrary type 1 degree of freedom.

Let us see where this ambiguity arises in an approach based upon chasing long exact sequences in cohomology. Consider the following sequence.
\begin{eqnarray} \label{captainseq}
H^1({\cal F}_1) \to H^1({\cal F}_0) \stackrel{f}{\longrightarrow} H^1(V_X) \stackrel{\delta}{\longrightarrow} H^2({\cal F}_1) \to H^2({\cal F}_0)
\end{eqnarray}
Type 1 fields are well defined. They are simply associated to elements of $H^1(V_X)$ which are in the image of $f$. One would like to define type 2 fields as those descending from $H^2({\cal F}_1)$, that is as elements of $H^1(V_X)$ that are in the preimage of $\delta$. This is ambiguous however. The exactness of the sequence (\ref{captainseq}) implies that $\delta \circ f=0$. Therefore if $\nu$ is an element of the preimage under $\delta$ of some element of $H^2({\cal F}_1)$ then so is $\nu + f(n)$ for any $n \in H^1({\cal F}_0)$. That is we can add an arbitrary cohomology element associated to a type 1 field to one describing a type 2 field and get another type 2 degree of freedom. Clearly this is exactly the same ambiguity that we observed in terms of Godement resolutions. 

Fortunately, the ambiguity just described will not prevent us from deriving vanishing theorems which hold however one chooses to resolve this ambiguity. In deriving these theorems, however, it will be important to be aware of this issue.

\subsection{Cup products and vanishing theorems via homological algebra}

Having established how one would describe type in an algebraic setting, and having introduced a few basic technical tools, let us proceed to apply some of the standard discussion from the mathematics literature concerning how one would go about computing the triple product corresponding to the Yukawa coupling. In this subsection we will continue to focus on the case where the resolution (\ref{seq1}) is short exact. We will generalize away from this restriction in the next subsection.

It is possible to write the following commutative diagram where the vertical maps are quasi-isomorphisms.
\begin{eqnarray} \label{square1}
\begin{array}{ccc}
\textnormal{Tot}[V_X]^{\otimes 3} &\to &\wedge^3 [V_X] \\ 
\uparrow&&\uparrow \\
{\textnormal{Tot}[{\cal F}_1 \to {\cal F}_0]^{\otimes 3}} & \to &\wedge^3[ {\cal F}_1 \to {\cal F}_0]
\end{array}
\end{eqnarray}
This diagram deserves some explanation. In forming a grid of sequences involving tensor products of the objects in $[{\cal F}_1 \to {\cal F}_0 \to V_X]$ we must be cautious. In general, the objects in this resolution are sheaves, not vector bundles, so the tensor product operation does not preserve short exactness of the sequences involved. The tensor product operation is right exact, however (the relevant left derived functor being Tor). Thus we can construct what are called third quadrant bi-complexes such as the following.
\begin{eqnarray} \label{example}
\begin{array}{ccccccc}
&0&&0&&&\\
&\uparrow&&\uparrow&&&\\
\to &{\cal F}_1\otimes {\cal F}_0& \to &{\cal F}_0\otimes {\cal F}_0&\to &0 \\
&\uparrow&&\uparrow&&&\\
\to &{\cal F}_1\otimes {\cal F}_1& \to &{\cal F}_0\otimes {\cal F}_1& \to &0\\
&\uparrow&&\uparrow&&&
\end{array}
\end{eqnarray}
All of the rows and columns in this grid of sequences are exact, except in their final entry. Indeed we can construct a three dimensional grid of sequences analogously. There is an associated, possibly infinite, total sequence associated to such grids which terminates on the right (we say it is `bounded above'). From an appropriate definition of the maps in such a total complex, in terms of the maps appearing in the complexes in grids such as (\ref{example}), one can show that this total complex is exact, except at the last entry. It is this complex which is denoted by ${\textnormal{Tot}[{\cal F}_1 \to {\cal F}_0]^{\otimes 3}} $ 
in (\ref{square1}). There exists a quasi-isomorphism between this complex and $\textnormal{Tot}[V_X]^{\otimes 3}$, which is the left hand vertical map in (\ref{square1}). This can be straightforwardly proven by showing that the relevant mapping cone is acyclic, see \cite{weibel} for details. Finally, one can apply anti-symmetrization of an appropriate type to these objects to obtain the rest of the diagram (see for example \cite{Anderson:2009ge}).

In fact the right hand map of (\ref{square1}) resembles something which is used in the literature on heterotic compactifications routinely. Consider the exterior power sequence \cite{stacks} associated to (\ref{seq1})
\begin{eqnarray} \label{extpower}
\ldots \to {\cal F}_1 \otimes \wedge^2 {\cal F}_0 \to \wedge^3 {\cal F}_0 \to \wedge^3V_X \to 0 \; .
\end{eqnarray}
This sequence is somewhat different to that associated to short exact sequences of bundles, for example as used in describing the spectra of monad bundles (see \cite{Anderson:2008ex} for a review), in that it does not terminate on the left. Nevertheless, (\ref{extpower}) can be turned into a quasi-isomorphism of complexes in exactly the same way that we turned (\ref{seq1}) into (\ref{quasi1}). The resulting quasi-isomorphism is the right hand map in (\ref{square1}).

Given the diagram (\ref{square1}) we can produce the following commutative diagram of bi-complexes.
\begin{eqnarray} \label{square2}
\begin{array}{ccc}
\textnormal{tot}[G_{\bullet}(V_X)]^{\otimes 3} &\to &[{\cal I}_{\bullet}(\wedge^3 V_X)] \\ 
\uparrow&&\uparrow \\
{\textnormal{tot}[G_{\bullet}({\cal F}_1) \to G_{\bullet}({\cal F}_0)]^{\otimes 3}} & \to & [{\cal I}_{\bullet}(\wedge^3[{\cal F}_1 \to {\cal F}_0])]
\end{array}
\end{eqnarray}
Here the two quantities on the left are partial total complexes, hence the use of `$\textnormal{tot}$' rather than `$\textnormal{Tot}$'. By this we simply mean that $(\textnormal{tot}[G_{\bullet}(V_X)]^{\otimes 3})_{i}=\bigoplus_{i=j+l+l} G_j(V_X) \otimes G_k(V_X)\otimes G_l(V_X)$ where the rest of the top left bi-complex is filled out with zeros and in ${\textnormal{tot}[G_{\bullet}({\cal F}_1) \to G_{\bullet}({\cal F}_0)]^{\otimes 3}} $ we take the total complex on the $G$ indices and ${\cal F}$ indices separately. That we can write the left hand side of the square in (\ref{square2}) is guaranteed by a property called `Universal Exactness' of the Godement resolution (see \cite{Anderson:2009ge} for a description appearing in the physics literature). The remaining three maps have their existence guaranteed by the existence of (\ref{square1}) and the properties of injective resolutions \cite{brylinski}. 

The diagram (\ref{square2}) is the algebraic structure associated to computing the Yukawa coupling. The top right hand corner is an injective resolution of $\wedge^3 V_X= {\cal O}_X$. The homology of global sections of this resolution is the definition of the sheaf cohomology of the trivial bundle. Thus the third such entry is the copy of $\mathbb{C}$ in which the Yukawa coupling between a given set of matter fields, as in (\ref{mrmappy}), resides. The top left of (\ref{square2}) is a combination of three copies of Godement resolutions of $V_X$. Thus it describes (upon considering homologies of global sections of the sheaves involved) a combination of three copies of the cohomology $V_X$. The top map in the diagram then contains the structure necessary to compute the Yukawa coupling, that is it contains all of the necessary information about the map (\ref{mrmappy}). The remainder of the commutative diagram (\ref{square2}) describes the structure that is imposed on this map by the existence of the resolution (\ref{seq1}). Roughly speaking, the left hand side of the diagram describes how the cohomology of $V_X$ is described in terms of cohomology of the sheaves ${\cal F}$, that is in terms of ambient space objects in the case that we consider a Koszul resolution such as (\ref{kos1}). It essentially encodes the sequence chasing of the long exact sequence associated to (\ref{seq1}), or (\ref{kos1}) in the Koszul special case. This is mapped along the bottom of the diagram to the anti-symmetrized product of the objects involved, which  then maps to the Yukawa coupling in the top right again due to the exactness of the exterior power sequence and the properties of the resolutions involved.

\vspace{0.1cm}

Having gained (\ref{square2}) we now wish to use it to understand where vanishing theorems such as \vref{v-theorem1} come from using a purely algebraic point of view. To see this it is instructive to write out the right hand side map of (\ref{square2}) a little more explicitly. We have 
\begin{eqnarray}  \label{fred}
\begin{array}{ccccccc}
\ldots  & \to & 0 & \to & {\cal I}_{\bullet}(\wedge^3 V_X) & \to & 0 \\
&& \uparrow && \uparrow &&\\
\ldots &  \to  & {\cal I}_{\bullet}({\cal F}_1 \otimes \wedge^2 {\cal F}_0) & \to & {\cal I}_{\bullet}( \wedge^3 {\cal F}_0) & \to & 0
\end{array} 
\end{eqnarray}
What we see from this diagram is that all of the non-trivial maps into $ {\cal I}_{\bullet}(\wedge^3 V_X)$ come through $ {\cal I}_{\bullet}( \wedge^3 {\cal F}_0) $. Therefore, if one were to construct the maps explicitly between the total complexes of the bi-complexes appearing in (\ref{square2}), and from there determine the maps between the cohomologies of the objects involved, then these would all be constructed following such a path.

Consider computing a $(\textnormal{type 1})^3$ coupling in the language of Section \ref{arg1}. Any element of $H^1(V_X)$ will be associated with a section of $G_1({\cal F}_0)$ according to (\ref{quasi2}). As mentioned in the preceding subsection, a field of type 1 will be associated to a section that is in the cohomology of the sequence $\Gamma(G_{\bullet}({\cal F}_0))$ (that is, $\Gamma(G_{\bullet}(V))$ in the case that (\ref{seq1}) is taken to be the Koszul sequence (\ref{kos1})). It is not hard to show that taking a tensor product of three such elements and following them along the bottom map in the diagram (\ref{square2}) leads to an element $\alpha \in \Gamma({\cal I}_3( \wedge^3 {\cal F}_0))$ which is in the cohomology of the $\Gamma({\cal I}_{\bullet}(\wedge^3 {\cal F}_0))$ sequence\footnote{To see this one uses that the maps in the Godement resolution directions in the bottom left complex of (\ref{square2}) are given by a Liebniz type rule up to sign (see \cite{Anderson:2009ge} for details), together with the defining properties of a map of bi-complexes as applied to  the morphism in the lower row of (\ref{square2}).}.  In {\it some cases} this cohomology will vanish. For example, if we are using the Koszul sequence (\ref{kos1}), then $H^3(\wedge^3 {\cal F}_0)=H^3(\wedge^3V)=0$ because we are restricting ourselves to the case of $SU(3)$ bundles in this paper for simplicity. In such a case, the element $\alpha \in \Gamma({\cal I}_3( \wedge^3 {\cal F}_0))$ is the image of an element $\beta \in \Gamma({\cal I}_2( \wedge^3 {\cal F}_0))$. Using the commutativity of the following piece of the previous diagram (which still holds once we take global sections),
\begin{eqnarray}
\begin{array}{ccc} 
{\cal I}_2 (\wedge^3 {\cal F}_0) & \to & {\cal I}_2(\wedge^3V_X) \\ 
\downarrow && \downarrow \\ 
{\cal I}_3(\wedge^3 {\cal F}_0) & \to & {\cal I}_3(\wedge^3V_X)
\end{array}
\end{eqnarray}
we then see that the element which $\alpha$ maps to in $\Gamma({\cal I}_3(\wedge^3V_X))$ is the image of some element in $\Gamma({\cal I}_2(\wedge^3V_X))$. Thus any such global section in $\Gamma({\cal I}_3(\wedge^3V_X))$ which is a cohomology element is the zero element in cohomology. Thus the Yukawa coupling vanishes.

In the case of other couplings, we see no such restriction. For example, a $(\textnormal{type 2})^3$ coupling is formed, in part, by mapping sections of $G_1({\cal F}_0)$ (that is $G_1(V)$ in the Koszul case) which are not in the cohomology of $\Gamma(G_{\bullet}({\cal F}_0))$ (respectively $\Gamma(G_{\bullet}(V))$). This means that they do not map to an element $\alpha \in \Gamma({\cal I}_3( \wedge^3 {\cal F}_0))$ which is in the cohomology of the sequence  $\Gamma(I_{\bullet}(\wedge^3 {\cal F}_0))$ and the same logic as above does not hold.

It is useful to note how closely this argument mimics the less abstract discussion of Section \ref{arg1}. We have a notion of type with sections of $G_1({\cal F}_0)$ replacing the forms $\hat{\nu}$ of the more concrete argument. We then find a type selection rule on Yukawa couplings which is the exact analogue of (\ref{th1}) from the algebraic approach. The only additional piece of information that goes into either computation, beyond the existence of a resolution such as (\ref{seq1}), is the fact that $H^3(\wedge^3 {\cal F}_0)=0$. This is used explicitly above, and in the argument of Section \ref{arg1} it is used implicitly when it is stated that $H^3(\wedge^3 V_X)=H^4({\cal N}^{\vee})$.

\vspace{0.2cm}

Finally, therefore, by abstracting the algebraic structure of \vref{v-theorem1} we can now easily generalize its codimension one formulation as follows.

\begin{v-theorem} \label{v-theorem2}
Given any short exact sequence,
\begin{eqnarray} \label{res}
0 \to {\cal F}_1 \to {\cal F}_0 \to V_X \to 0\;,
\end{eqnarray}
we can define a notion of `type' for elements of $H^1(V_X)$. Namely, we say that such a cohomology element has type $\tau=1$ if it descends from an element of $H^1({\cal F}_0)$ and type $\tau=2$ if it descends from an element of $H^2({\cal F}_1)$. Then, if $H^3(\wedge^3 {\cal F}_0)=0$, a Yukawa coupling will vanish unless the sum of the types of the cohomology elements involved is greater than or equal to the length of the resolution (\ref{res}) plus three (in this case greater than or equal to four). Denoting the types of the matter fields of interest by $\tau_i$ with $i=1,\ldots,3$, and denoting the length of the resolution (\ref{res}) by $L$ ($=1$ in this case), we have the following
\begin{eqnarray} \label{gnth1a}
\tau_1 +\tau_2 + \tau_3 < L+3 \Rightarrow \lambda_{123} =0 \; .
\end{eqnarray}
\end{v-theorem}

In stating the above we have used conventions that make the relation between (\ref{gnth1a}) and (\ref{th1}) most apparent. The historical definition of type is, however, rather unfortunate given the structure we see in the current discussion. It would be more natural to label types $1$ and $2$ as $0$ and $1$ respectively. That is, we would label the type of a field associated to a cohomology element in $H^\bullet({\cal F}_i)$ as $i$. Then, (\ref{gnth1a}) would simply read as follows 
\begin{eqnarray} \label{gnth1b}
\tau_1 +\tau_2 + \tau_3 < L \Rightarrow \lambda_{123} =0 \; .
\end{eqnarray}

\vspace{0.2cm}

We will now generalize this argument beyond the case of length one resolutions, i.e. we will go beyond short exact sequences such as (\ref{res}). This will provide us with a generalization of the vanishing theorems obtained in \cite{Blesneag:2016yag} from higher codimension Koszul sequences. The essential core of the argument, however, will be unchanged from the structure we have stated above.

\subsection{Generalization to longer resolutions}

Let us now consider performing a similar analysis for the case where the resolution of $V_X$ can be of any length, not simply a short exact sequence as in (\ref{seq1}). In particular, this analysis will include as a special case Koszul resolutions of Calabi-Yau manifolds described as complete intersections of codimension two and higher. We start with the following left resolution
\begin{eqnarray} \label{seq2}
\ldots \to {\cal F}_2 \to {\cal F}_1 \to {\cal F}_0 \to V_X \to 0 \; .
\end{eqnarray}
 An example of such a resolution would be Koszul in codimension two 
\begin{eqnarray} \label{kos2}
 0 \to \wedge^2 {\cal N}^{\vee} \otimes V \to {\cal N}^{\vee} \otimes V \to V \to V_X \to 0 \; .
\end{eqnarray}
In such a case one would have that ${\cal F}_0=V$, ${\cal F}_1 = {\cal N}^{\vee} \otimes  V$, ${\cal F}_2 =\wedge^2 {\cal N}^{\vee} \otimes V$ and ${\cal F}_i=0$ for all $i \geq 3$.

Introducing suitable kernels/images (\ref{seq2}) can be split into a set of short exact sequences.
\begin{eqnarray} \label{kos22}
\begin{array}{ccccccccc}
 0 & \to &k_1 &\to &{\cal F}_0 &\to &V_X &\to& 0 \\
 0 & \to &k_2 &\to &{\cal F}_1  &\to &k_1 &\to &0 \\ 
  0 & \to & k_3 & \to &{\cal F}_2  &\to &k_2& \to &0 \\ 
&&&& \vdots &&&&
\end{array}
\end{eqnarray}

Given that the first sequence in (\ref{kos22}) is short exact we can reproduce much of the analysis of the previous subsection with the simple replacement of ${\cal F}_1$ with $k_1$. In particular, the vanishing theorem concerning couplings corresponding to three matter fields all of type 1, that is all corresponding to an element of $H^1({\cal F}_0)$, still holds. However, some extra structure can be observed in addition, corresponding to the vanishing that is found in \cite{Blesneag:2016yag} for couplings involving higher types, once one considers codimensions above one.

In the situation at hand, the diagram (\ref{square2}) is replaced with
\begin{eqnarray} \label{square3} 
\begin{array}{ccc}
\textnormal{tot}[G_{\bullet}(V_X)]^{\otimes 3} &\to &[{\cal I}_{\bullet}(\wedge^3 V_X)] \\ 
\uparrow&&\uparrow \\
{\textnormal{tot}[G_{\bullet}(k_1)\to G_{\bullet}({\cal F}_0)]^{\otimes 3}} & \to & [{\cal I}_{\bullet}(\wedge^3[k_1 \to {\cal F}_0])]
\end{array}
\end{eqnarray}
The first sequence in (\ref{kos22}) has an associated exterior power sequence \cite{stacks} that is exact on the right - and it is essentially an injective resolution of this that forms the right hand side of (\ref{square3}). Note also that from the second sequence in (\ref{kos22}) we can write,
\begin{eqnarray} \label{feedthrough1}
\begin{array}{ccccccc}
0&\to& 0&\to&    G_{\bullet}(k_1)& \to& 0\\
&&\uparrow&&\uparrow&&\\
0&\to& G_{\bullet}(k_2)& \to & G_{\bullet}({\cal F}_1)&  \to&0
\end{array}
\end{eqnarray}
which gives a quasi-isomorphism at each level in the Godement resolutions. Note that this set of sequences is analogous to (\ref{quasi2}) in Subsection \ref{seqchase}.  This allows us to follow elements of $G_2({\cal F}_1)$ through to $G_2(k_1)$ and from there into the bottom left of (\ref{square3}).

\vspace{0.1cm}

We will now use the above to study the structure of couplings between one type 2 and two type 1 fields. The type 1 degrees of freedom will be associated to two global sections, $\nu_1$ and $\nu_2$ of $G_1({\cal F}_0)$ which are closed under the Godement map. The type 2 field is associated to a Godement closed section $w \in \Gamma(G_2({ {\cal F}_1}))$ and a non-closed section $v \in \Gamma(G_1({\cal F}_0))$. As discussed in Section \ref{seqchase}, these two are related by $dv=f(w)$ where $d$ is the Godement map and $f$ is a map associated to the resolution (\ref{seq2}). The global section $w$ is associated, via (\ref{feedthrough1}), with a global section $\tilde{w}\in \Gamma(G_2(k_1))$. It is straightforward to show, using the commutativity of the following piece of (\ref{feedthrough1}), 
\begin{eqnarray}
\begin{array}{ccc} 
{G}_2 ( {\cal F}_1) & \to & {G}_2(k_1) \\ 
\downarrow && \downarrow  \\
{G}_3({\cal F}_1) & \to & {G}_3(k_1) 
\end{array}
\end{eqnarray}
which also holds at the level of global sections, that if $w$ is closed under the Godement map then so is $\tilde{w}$.

The sections of Godement sheaves described in the previous paragraph combine to form global sections $A \in \Gamma(G_1({\cal F}_0) \otimes G_1({\cal F}_0)\otimes G_2(k_1))$ and $B \in \Gamma(G_1({\cal F}_0) \otimes G_1({\cal F}_0) \otimes G_1({\cal F}_0))$ which appear inside certain objects in the bi-complex in the bottom left of (\ref{square3}). Following the properties of the induced maps in that bi-complex these global sections obey $dB =f(A)$ where $d$ is the induced Godement and $f$  the other, horizontal, map (where by a slight abuse of notation we use the same symbols for these as for the maps from which they are derived).

Denoting the morphism of bi-complexes along the bottom of (\ref{square3}) by $\phi$, the global sections $A$ and $B$ map to $B'=\phi(B) \in \Gamma( {\cal I}_3(\wedge^3 {\cal F}_0))$ and $A'=\phi(A) \in  \Gamma({\cal I}_4(k_1 \otimes \wedge^2 {\cal F}_0))$  which appear inside certain objects in the bi-complex on the bottom right of (\ref{square3}). To be a morphism of bi-complexes, the map $\phi$ obeys two intertwining conditions, which in our language can be written as $d' \phi= \phi d$ and $f' \phi = \phi f$. Here the morphism $d'$ is the morphism associated to the injective resolutions and the morphism $f'$ is the other, horizontal, set of maps. Using this one can easily prove that $d'B'=f'A'$. 

Consider now a situation where $H^4({\cal F}_1 \otimes \wedge^2 {\cal F}_0)=0$. Given the universal exactness of Godement resolutions, our global sections $\nu_1$, $\nu_2$ and $w$ can be mapped to a global section $\tilde{A} \in \Gamma({\cal I}_4({\cal F}_1 \otimes \wedge^2 {\cal F}_0))$ which is closed under the map associated to the injective resolution. By tensoring up the middle sequence in (\ref{kos22}) by $\wedge^2 {\cal F}_0$ and taking injective resolutions we can write the following commutative diagram.
\begin{eqnarray}
\begin{array}{ccc} 
{\cal I}_3 ( {\cal F}_1 \otimes \wedge^2 {\cal F}_0) & \to & {\cal I}_3(k_1 \otimes \wedge^2 {\cal F}_0)  \\ 
\downarrow && \downarrow  \\
{\cal I}_4({\cal F}_1 \otimes \wedge^2 {\cal F}_0) & \to & {\cal I}_4(k_1 \otimes \wedge^2 {\cal F}_0)  
\end{array}
\end{eqnarray}
Our section $A'$ of the proceeding paragraph is the image under the lower map in this diagram of $\tilde{A}$. The vanishing of $H^4({\cal F}_1 \otimes \wedge^2 {\cal F}_0)$ ensures that $\tilde{A}$ is the image of some element $\tilde{C}\in \Gamma({\cal I}_3 ( {\cal F}_1 \otimes \wedge^2 {\cal F}_0) )$. Commutativity of this diagram immediately shows that any section in the image of the vertical map on the left maps to a section in the image of the vertical map on the right. Thus our section $A'$ is the image of some element $C' \in \Gamma ( {\cal I}_3(k_1 \otimes \wedge^2 {\cal F}_0))$. 

\vspace{0.1cm}

We now have all of the information we require to compute the Yukawa coupling between these fields. The exterior power sequence \cite{stacks} that is used to form the right hand side of (\ref{square3}) can be split into pieces by appropriate use of kernels and cokernels. The first two such short exact pieces are as follows.
\begin{eqnarray} \label{splitext}
&&0\to \alpha_1 \to \wedge^3 {\cal F}_0 \to \wedge^3 V_X \to0\\ \nonumber
&&0\to \alpha_2 \to k_1 \otimes \wedge^2 {\cal F}_0 \to \alpha_1 \to 0
\end{eqnarray}
This leads to the following pieces of grids of sequences describing some of the objects that appear in the right hand side of (\ref{square3}).
\begin{eqnarray} \label{oneone}
\begin{array}{ccccccc}
{\cal I}_3(\alpha_1)&\stackrel{f'}{\to}&{\cal I}_3(\wedge^3 {\cal F}_0) &\stackrel{g'}{\to}& {\cal I}_3(\wedge^3 V_X)& \to& 0\\
\downarrow &&\downarrow &&\downarrow &&\\
{\cal I}_4(\alpha_1)&\to&{\cal I}_4(\wedge^3 {\cal F}_0) &\to& {\cal I}_4(\wedge^3 V_X)& & \end{array}
\end{eqnarray}
\begin{eqnarray} \label{twotwo}
\begin{array}{ccccccc}
&&{\cal I}_3(k_1\otimes \wedge^2 {\cal F}_0) &\stackrel{h}{\to}& {\cal I}_3(\alpha_1)& \to& 0\\
&&\downarrow &&\downarrow&&\\
{\cal I}_4(\alpha_2)&\to&{\cal I}_4(k_1\otimes \wedge^2 {\cal F}_0) &\to& {\cal I}_4(\alpha_1)& &
\end{array}
\end{eqnarray}
Given the discussion up to this point, we have an element $A' \in\Gamma({\cal I}_4(k_1 \otimes \wedge^2 {\cal F}_0))$ which is the image of an element $C' \in \Gamma( {\cal I}_3(k_1 \otimes \wedge^2 {\cal F}_0))$ and a Godement closed element $B' \in \Gamma( {\cal I}_3(\wedge^3 {\cal F}_0))$ which we wish to map through the diagram to obtain the Yukawa coupling which is given by a global section of $ {\cal I}_3(\wedge^3 V_X)$.

To map $A'$ through the diagram to $\Gamma( {\cal I}_3(\wedge^3 V_X))$ we can take its preimage $C' \in \Gamma( {\cal I}_3(k_1 \otimes \wedge^2 {\cal F}_0))$ and map this to $h(C') \in \Gamma({\cal I}_3(\alpha_1))$. One can then form $g'(f'(h(C'))) \in \Gamma({\cal I}_3(\wedge^3 V_X))$, following along the top line of the diagram (\ref{oneone}).  Now the exactness of the top line of that diagram implies that $g'\circ f' =0$ and this extends to global sections. Therefore $A'$ maps through to a vanishing contribution in $\Gamma( {\cal I}_3(\wedge^3 V_X))$. Mapping a different route through the diagrams will lead to the same element in $\Gamma({\cal I}_3(\wedge^3 {\cal F}_0))$ up to a piece that is closed under the map of the injective resolution. 

Mapping $B' \in \Gamma({\cal I}_3(\wedge^3 {\cal F}_0))$ through the diagram is much simpler. By commutativity of all of the diagrams involved and the fact that $dB' =f(A')$ (where $f$ is actually a combination of maps in the above diagrams) we have that $B'$ is equal to $f'(h(C'))$ up to a closed piece. This possible closed piece is exactly the same as that described in the previous paragraph. Thus, we need only consider a closed element of $\Gamma({\cal I}_3(\wedge^3 {\cal F}_0))$ and how it would contribute to the Yukawa couplings. If we impose the additional condition that $H^3(\wedge^3 {\cal F}_0)=0$ then the same arguments as those given in the previous subsection tell us that this contribution to the Yukawa coupling will vanish as well.

\vspace{0.1cm}

Returning to the special case of the Koszul resolution, the vanishing $H^4({\cal F}_1 \otimes \wedge^2{\cal F}_0)=0$ becomes $H^4({\cal N}^{\vee} \otimes V \otimes \wedge^2 V)=0$. Such a vanishing would not hold, in general, in the higher codimension cases considered in \cite{Blesneag:2016yag}. To see what happens there consider that $H^4({\cal N}^{\vee}\otimes V \otimes \wedge^2 V)=H^4({\cal N}^{\vee} \otimes V \otimes V^{\vee})=H^4({\cal N}^{\vee}) \oplus H^4({\cal N}^{\vee} \otimes \textnormal{End}_0(V))$, where we have used the fact that $V$ is an $SU(3)$ bundle in that case. In the case of a complete intersection Calabi-Yau $H^4({\cal N}^{\vee}) =0$, that is {\it part} of the cohomology $H^4({\cal N}^{\vee}\otimes V \otimes \wedge^2 V)$ 
vanishes. Because of the anti-symmetrization structure in the maps in (\ref{square3}), this is enough to secure the vanishing Yukawa coupling. 
The direct generalization of the types of vanishing theorems seen in the Koszul complex at codimension two and beyond \cite{Blesneag:2016yag} is thus somewhat baroque. We have the following.

\begin{v-theorem} \label{v-theorem3}  Given any  exact sequence,
\begin{eqnarray} \label{res2}
\ldots \to {\cal F}_2 \to {\cal F}_1 \to {\cal F}_0 \to V_X \to 0\;,
\end{eqnarray}
 we can define a notion of `type' for elements of $H^1(V_X)$. Namely, we say that such a cohomology element has type $\tau=i$ if it descends from an element of $H^{i}({\cal F}_{i-1})$. Then, if $H^3(\wedge^3 {\cal F}_0)=0$, all Yukawa couplings between three type $\tau=1$ fields will vanish. If, in addition, ${\cal F}_1 = U \otimes {\cal F}_0$ for some $U$ and if $H^4(U \otimes \wedge^3 {\cal F}_0)=0$ then all Yukawa couplings between one type $\tau=2$ and two type $\tau=1$ fields will vanish.
\end{v-theorem}

If the resolution (\ref{res2}) is of length $2$ and both of the above cohomological vanishings hold, then a Yukawa coupling will vanish unless the sum of the types of the cohomology elements involved is greater than or equal to the length of the resolution (\ref{res2}) plus three (in this case greater than or equal to five). Denoting the types of the matter fields of interest by $\tau_i$ with $i=1,\ldots,3$, and denoting the length of the resolution (\ref{res2}) by $L$ ($=2$ in this case), we have
\begin{eqnarray} \label{gnth2a}
\tau_1 +\tau_2 + \tau_3 < L+3 \Rightarrow \lambda_{123} =0 \; .
\end{eqnarray}
This is the direct analogue of the vanishing theorem given for the codimension two Koszul complex in \cite{Blesneag:2016yag}. As with the previous vanishing theorem, the above conventions make the relation between (\ref{gnth2a}) and (\ref{th1}) most apparent. However, it would be more natural in the general context to change the definition of type by subtracting unity.
\vspace{0.2cm}

It is clear from the discussion of this section that we could have a less complicated vanishing result that applied to a less restrictive set of resolutions, at the expense of requiring a stronger cohomological vanishing condition than that appearing in \vref{v-theorem3}.

\begin{v-theorem} \label{v-theorem4}
Given any exact sequence,
\begin{eqnarray} \label{res3}
\ldots \to {\cal F}_2 \to {\cal F}_1 \to {\cal F}_0 \to V_X \to 0\;,
\end{eqnarray}
 we can define a notion of `type' for elements of $H^1(V_X)$. Namely, we say that such a cohomology element has type $\tau=i$ if it descends from an element of $H^{i}({\cal F}_{i-1})$. Then, if $H^3(\wedge^3 {\cal F}_0)=0$, all Yukawa couplings between three type $\tau=1$ fields will vanish. If, in addition, $H^4({\cal F}_1 \otimes \wedge^2 {\cal F}_0)=0$ then all Yukawa couplings between one type $\tau=2$ and two type $\tau=1$ fields will also vanish.
\end{v-theorem}

It is important to stress at this stage that, with the simple analysis being carried out here, we can not go further and analyze the analogue of vanishings of couplings between even higher type fields seen in \cite{Blesneag:2016yag}. The reason for this is that exterior power sequences such as (\ref{extpower}) associated to (\ref{seq2}), as presented in decomposed form in (\ref{splitext}), start to become more complicated at the next entry beyond the ones explicitly shown in those formulae. Tor functors start to appear in the relevant entries in the sequences and, while it would be interesting to analyze this structure further, such work is beyond the scope of this paper.

\vspace{0.2cm}

As one final point in our general discussion we would like to highlight that the resolution (\ref{res3}) must be over some space whose dimension is greater than that of $X$ if the vanishing theorems are to be useful (as is the case in the Koszul resolution for example). The reason for this is that the exterior power sequence associated to (\ref{res3}) ends as follows.
\begin{eqnarray}
\wedge^3 {\cal F}_0 \to \wedge^3 V_X \to 0
\end{eqnarray}
If the variety over which this sequence is being written is three dimensional then cohomology groups are concentrated in degrees less than or equal three and this leads to the following fragment of the long exact sequence
\begin{eqnarray}
H^3(\wedge^3 {\cal F}_0) \to H^3(\wedge^3 V_X) \to 0 \; .
\end{eqnarray}
The bundle $V_X$ has structure group $SU(3)$ and thus $H^3(\wedge^3 V_X)=H^3({\cal O}_X) = \mathbb{C}$. Thus, in such a situation $H^3(\wedge^3{\cal F}_0)$ can never vanish and so none of the theorems we have discussed hold.

Although the resolution (\ref{res3}) can not be one over $X$, it is hopefully clear from the analysis of this section that it does not have to be the Koszul resolution. However, even restricting attention to the Koszul case, the same bundle $V_X$ admits a plethora of different resolutions of that type, involving both different ambient spaces for $X$ and different sheafy lifts of $V_X$. This can frequently lead to a variety of different patterns of vanishing Yukawa couplings enforced by the different possible choices for (\ref{res3}). We will illustrate this rather ubiquitous phenomenon with concrete examples in the next section.

%%%%%%%%%%%%%%%%%%%%%%%%

\section{Examples} \label{secegs}

\subsection{Ambiently defined monad bundles}

A simple case in which the theorems of the previous section can be applied is that of ambiently defined bundles. This is a very simple generalization of the results presented in \cite{Blesneag:2015pvz,Blesneag:2016yag}. Consider an $SU(3)$ bundle $V_X$ on a Calabi-Yau $X$ which is the restriction of an $SU(3)$ bundle $V$ on an ambient space ${\cal A}$ of one dimension higher. Further, consider a case where $h^3({\cal O})=0$ on the ambient space. In such a situation we have that $h^3(\wedge^3 V)=h^3({\cal O})=0$ and thus Vanishing Theorem \ref{v-theorem3}, as applied to the Koszul resolution,
\begin{eqnarray} \label{kosagain}
0\to {\cal N}^{\vee} \otimes V\to V \to V_X \to 0
\end{eqnarray}
says that  all couplings between three fields descending from elements of $H^1(V)$ will vanish. 

\vspace{0.2cm}

To give a concrete example of this, consider the following monad bundle
\begin{eqnarray} \label{mon1}
0 \to V_X \to {\cal O}_X(1,1)^{\oplus 3} \oplus {\cal O}_X(3,-2) \to {\cal O}_X(6,1) \to 0 \;,
\end{eqnarray}
over the CICY \cite{Yau:1986gu,Hubsch:1986ny,Candelas:1987kf,Candelas:1987du,Green:1986ck,Gray:2013mja,Gray:2014fla,Anderson:2015iia,Berglund:2016yqo,Berglund:2016nvh} defined by the following configuration matrix
\begin{eqnarray} \label{24one}
X = \left[ \begin{array}{c|c} \mathbb{P}^1 &2 \\ \mathbb{P}^3 &4\end{array}\right] \; .
\end{eqnarray}
The matrix (\ref{24one}) simply indicates that $X$ is defined as the vanishing locus of a degree $(2,4)$ polynomial in $\mathbb{P}^1 \times \mathbb{P}^3$. The bundle $V_X$ is the restriction of a similar object on the ambient space
\begin{eqnarray} \label{mon1amb} 
0 \to V \to {\cal O}(1,1)^{\oplus 3} \oplus {\cal O}(3,-2) \stackrel{f}{\longrightarrow} {\cal O}(6,1) \to 0 \; .
\end{eqnarray}
Note in particular that the first three entries in the map $f$ here  would be a set of 3 quintics in $\mathbb{P}^1$. As such the rank of a sufficiently general $f$ is maximal everywhere on $\mathbb{P}^1 \times \mathbb{P}^3$ and thus $V$ is a bundle. A straightforward computation using the usual techniques (see for example \cite{Anderson:2009nt} for a brief description) reveals that $V_X$ is stable below the line of slope 4 in the K\"ahler cone.

Finally, one can easily show that
\begin{eqnarray}
h^1(V_X) = 4  \;\;\textnormal{and} \;\; h^2({\cal N}^{\vee} \otimes V) =0\;.
\end{eqnarray}
Thus all of the families in this example are of type 1 with respect to the sequence (\ref{kosagain}) and thus all of their superpotential Yukawa couplings in the associated four dimensional effective theory vanish. 

\vspace{0.1cm}

This example was chosen specifically because this result can be verified by completely distinct means. In studying the origins of the generations (as opposed to anti-generations) in this example, the line bundle ${\cal O}_X(3,-2)$ in (\ref{mon1}) plays no role. Thus (\ref{mon1}) acts much like a positive monad in terms of computing the Yukawa couplings between the families. As such, the methods espoused in \cite{Distler:1995bc,Anderson:2009ge} can be employed and we find that the couplings between families do indeed vanish as expected. Note that the methodology appearing in \cite{Anderson:2009ge} uses similar technology to that seen in the last section. However, in that work only sequences of bundles are studied. In particular the defining sequence of the monad and its long exact exterior power sequence plays the central role. As such those techniques can not be directly applied to a sequence of sheaves such as (\ref{kosagain}). 

\subsection{Multiple constraints from different ambient spaces}

In this section we wish to consider an example that demonstrates that a single bundle can obtain different constraints from different possible resolutions. In order to keep this example as simple as possible we will study a case which could have easily been analyzed using the techniques of \cite{Blesneag:2015pvz,Blesneag:2016yag}: a sum of line bundles descending directly from the ambient space. By considering different ambient spaces we will then obtain different constraints on the same $V_X$. In the next section, we will show that an array of different constraints can also be obtained for different resolutions associated to different sheaves on the same ambient space that restrict to the same bundle $V_X$ on the Calabi-Yau. For those examples we will need the flexibility of our algebraic analysis.

Consider the bundle
\begin{eqnarray} \label{mrbun1}
V_X = {\cal O}_X(-1,-1,1)^{\oplus 2} \oplus {\cal O}_X(2,2,-2)
\end{eqnarray}
over the CICY,
\begin{eqnarray} \label{x1}
X= \left[\begin{array}{c|cc} \mathbb{P}^1 & 1 & 1 \\ \mathbb{P}^1&1&1 \\ \mathbb{P}^3&4&0\end{array}\right] \;.
\end{eqnarray}
This bundle has 8 generations and 30 anti-generations. With respect to the long exact Koszul resolution associated to the above construction all of the generations are of type 2 descending from $H^2({\cal O}(-2,-2,1)^{\oplus 2}) \subset H^2({\cal N}^{\vee} \otimes V_X)$. Thus, with this description of $V_X$, one can  not use the vanishing theorems described in the previous sections to say anything about the structure of the couplings. All Yukawa couplings of three families are between three type 2 degrees of freedom.

However, the same manifold can, of course, be embedded in different ambient spaces. Consider the following configuration matrix.
\begin{eqnarray} \label{24}
X= \left[ \begin{array}{c|c} \mathbb{P}^1 & 2\\ \mathbb{P}^3 & 4\end{array}\right]
\end{eqnarray}
This manifold is in fact the same as that described by (\ref{x1}). There are several different ways of observing this equivalence. One is to simply note that (\ref{x1}) is an ineffective split of (\ref{24}) \cite{Candelas:1987kf}. Another is to observe that 
\begin{eqnarray} \label{p1}
\mathbb{P}^1 = \left[ \begin{array}{c|c} \mathbb{P}^1 & 1 \\ \mathbb{P}^1 & 1\end{array}\right] \;,
\end{eqnarray}
which can then be used to argue the equivalence of (\ref{24}) and (\ref{x1}) \cite{Candelas:1987kf}. 

In the description (\ref{p1}) of $\mathbb{P}^1$ the line bundle ${\cal O}_{\mathbb{P}^1} (a,b)$ can be shown to be the $(a+b)$'th power of the hyperplane line bundle, or ${\cal O}(a+b)$ in the more standard description. Given this equivalence, the line bundle ${\cal O}_X(a,b,c)$ on (\ref{x1}) is the same bundle as ${\cal O}_X(a+b,c)$ on (\ref{24}). Thus, describing $X$ via (\ref{24}) the bundle (\ref{mrbun1}) corresponds to the following.
\begin{eqnarray}
V_X = {\cal O}_X(-2,1)^{\oplus 2} \oplus  {\cal O}_X(4,-2)
\end{eqnarray}
In terms of the Koszul sequence associated to (\ref{24}), the eight generations of $V_X$ in this description are all of type 1, descending from $H^1({\cal O}(-2,1)^{\oplus 2})\subset H^1(V)$. Thus the vanishing theorems we have seen in Sections \ref{arg1} and \ref{arg2} tell us that all of the Yukawa couplings between families in this compactification vanish. 

The above is an explicit demonstration that the vanishing theorems we have been discussing, applied to different resolutions of the same bundle, lead to different restrictions on the Yukawa couplings. In this example the Koszul sequence associated to (\ref{x1}) gives us no information whereas that associated to (\ref{24}) reveals that all of the Yukawa couplings vanish. 

The above example was chosen for two reasons. First, this result can be confirmed by examining the constraints due to certain $U(1)$ symmetries that are present in the four dimensional theory associated to the bundle $V_X$. The bundle (\ref{mrbun1}) is an $S(U(1) \times U(1) \times U(1))$ bundle rather than a true $SU(3)$ object. As such, the four dimensional effective theory to which it is associated will have two $U(1)$ symmetries, in addition to the usual $E_6$ commutant of $SU(3)$ inside $E_8$. It is a simple matter to show that these $U(1)$ symmetries do indeed forbid all couplings between families in this example - in agreement with the above result from the vanishing theorems (see \cite{Blumenhagen:2005ga,Blumenhagen:2006ux,Blumenhagen:2006wj,Anderson:2011ns,Anderson:2012yf} for a few examples of such analyses). 

The second reason for chosing this example is that it can be used as a precursor case for the next source of multiple resolutions that we will discuss.
Because (\ref{x1}) is hyperfavorable, having a number of ambient space factors larger than $h^{1,1}$, many different ambient space bundles can restrict to the same object on the Calabi-Yau. Given the correspondence between line bundles on (\ref{24}) and (\ref{x1}) discussed above, $V_X$ on (\ref{x1}) could equally well be described as follows, rather than as was presented in (\ref{mrbun1}).
\begin{eqnarray} \label{bunbun2}
V_X = {\cal O}_X(0,-2,1)^{\oplus 2} \oplus {\cal O}_X (0,4,-2)
\end{eqnarray}
In the description (\ref{bunbun2}) all of the families are of type 1, descending from $H^1({\cal O}(0,-2,1)^{\oplus 2})\subset H^1(V)$. Thus the constraints one would get from the vanishing theorems discussed in this paper, actually depend on not only the ambient space that is being discussed, but also the lift of the bundle $V_X$ to that ambient space. In what follows we will give a more complex example of this phenomenon where some of the lifts of $V_X$ are sheaves on the ambient space rather than bundles. For this we will need the more general framework for vanishing theorems provided by the algebraic approach of Section \ref{arg2}.

\subsection{Multiple constraints from different sheafy lifts of the bundle}

Let us look at some examples where a bundle on $X$ descends from a sheaf on an ambient space ${\cal A}$. We will consider the following Calabi-Yau threefold,
\begin{eqnarray} 
X= \left[\begin{array}{c|c} \mathbb{P}^1 & 2 \\ \mathbb{P}^3& 4 \end{array} \right]\;,
\end{eqnarray}
together with the simple sum of line bundles,
\begin{eqnarray} \label{sum1}
V_X= {\cal O}_X(1,-2)^{\oplus 2} \oplus {\cal O}_X(-2,4) \;.
\end{eqnarray}
This bundle has $h^1(V_X)=32$ with all $32$ generations arising from $H^1({\cal O}_X(-2,4))$

Naturally, (\ref{sum1}) can be viewed as the restriction of an analogous sum of line bundles on the ambient space. However, (\ref{sum1}) can also be obtained by the restrictions of certain sheaves. For example, let us specify four points in $\mathbb{P}^1 \times \mathbb{P}^3$, via the vanishing locus of four degree $(1,1)$ polynomials. Let us define the ideal sheaf associated to these points as ${\cal I}_4$. We then have the following defining sequence for that sheaf.
\begin{eqnarray} \label{idealdef1}
0\to {\cal I}_4 \to {\cal O} \to {\cal O}_4 \to 0
\end{eqnarray}
Here ${\cal O}_4$ is the trivial bundle restricted to the four points. By comparing (\ref{idealdef1}) to the Koszul resolution for the four points we find that we have another exact sequence describing ${\cal I}_4$.
\begin{eqnarray} \label{idealdef2}
0 \to \wedge^4 {\cal N}_4^{\vee} \to \ldots \to {\cal N}_4^{\vee} \to {\cal I}_4 \to 0
\end{eqnarray}
In this expression ${\cal N}_4= {\cal O}(1,1)^{\oplus 4}$ is the normal bundle to the four points.

With these preparations, we can now define a sheaf on ${\cal A}$ that restricts to $V_X$ on $X$
\begin{eqnarray} \label{sheafyraise1}
V= {\cal O}(1,-2)^{\oplus 2} \oplus {\cal O}(-2,4) \otimes {\cal I}_4 \; . 
\end{eqnarray}
The essential point is that the points where the rank of the stalks of the sheaf ${\cal O}(-2,4) \otimes {\cal I}_4$ degenerate miss the Calabi-Yau. Therefore  ${\cal O}(-2,4) \otimes {\cal I}_4$ restricts to ${\cal O}_X(-2,4)$ as can be verified by restricting the sequence (\ref{idealdef1}) to the Calabi-Yau. 

Despite the fact that $V$ is a sheaf, not a vector bundle, we can write a short exact Koszul sequence in this case of the usual form
\begin{eqnarray} \label{kosyetagain}
0 \to {\cal N}^{\vee} \otimes V \to V \to V_X \to 0 \; .
\end{eqnarray}
The reason for this is that the exactness or not of a sequence of sheaves is a local issue. It can, therefore be checked using a fine enough open cover. Choose an open cover such that no element intersects both the four points associated to the ideal sheaf and the locus of the Calabi-Yau threefold $X$. If we consider an open set of our cover that intersects $X$ then $V$ is locally free on this set and we can tensor the sequence
\begin{eqnarray} \label{purekos}
0 \to {\cal N}^{\vee} \otimes {\cal O} \to {\cal O} \to {\cal O}_X \to 0
\end{eqnarray}
 by $V$ while maintaining exactness. Conversely, if we consider an open set which intersects one of the four points associated to ${\cal I}_4$ then all of the bundles in the sequence (\ref{purekos}) are locally free on this set. Thus we can tensor that sequence by $V$ while maintaining exactness.

By using combinations of long exact sequences in cohomology associated to (\ref{idealdef1}) and (\ref{idealdef2}) (once the latter is broken up into short exact pieces) tensored by appropriate line bundles, one can compute\footnote{To expedite and check some of the longer computations in this section the authors  made use of the CICY package \cite{CICYpackage}.}
\begin{eqnarray}
h^*({\cal N}^{\vee} \otimes {\cal O}(-2,4) \otimes {\cal I}_4 )&=& (0,7,0,0,0) \\ \nonumber
h^*( {\cal O}(-2,4) \otimes {\cal I}_4 ) &=& (0,39,0,0,0) \; .
\end{eqnarray}
Using these values in the long exact sequence in cohomology associated to (\ref{kosyetagain}) we find that all of the families in this example are of type 1.
From (\ref{sheafyraise1}) we can easily see that $\wedge^3 V = {\cal I}_4$. From the long exact sequence in cohomology associated to (\ref{idealdef1}) we therefore see that $h^3(\wedge^3 V) =h^3( {\cal I}_4) =0$. Thus, from \vref{v-theorem4}, we find that all Yukawa couplings between the families vanish in this example.

As in the previous subsection, this example has been chosen in part because this answer can be verified. The bundle (\ref{sum1}) is an $S(U(1) \times U(1) \times U(1))$ bundle rather than a true $SU(3)$ object and, as such, the four dimensional effective theory to which it is associated will again have two $U(1)$ symmetries. It is a simple matter to show that these $U(1)$ symmetries do indeed forbid all couplings between families in this example.

\vspace{0.2cm}

In the example we just looked at we obtain the same results if we lift (\ref{sum1}) to the ambient space as the obvious simple sum of line bundles rather than the more exotic sheafy object (\ref{sheafyraise1}). Therefore, to illustrate the importance of considering all possible sheaf and bundle lifts of $V_X$ to the ambient space, we now give a slightly more complex example where such a coincidence of constraints does not occur.

\vspace{0.2cm}

Consider defining, on the Calabi-Yau manifold given by the configuration matrix
\begin{eqnarray}
X=\left[ \begin{array}{c|cc} \mathbb{P}^1 &1 &1 \\\mathbb{P}^1 &1 &1 \\\mathbb{P}^1 &1 &1 \\\mathbb{P}^1 &1 &1 \\\mathbb{P}^1 &1 &1  \end{array} \right]\;,
\end{eqnarray}
the following sum of line bundles.
\begin{eqnarray} \label{VX2}
V_X= {\cal O}_X(2,-2,0,0,0) \oplus {\cal O}_X(-1,1,0,0,0)^{\oplus 2}
\end{eqnarray}
This bundle has the following cohomology,
\begin{eqnarray}
h^*(V_X) =(0,3,3,0) \;,
\end{eqnarray}
with all of the non-zero entries coming from the cohomology of ${\cal O}_X(2,-2,0,0,0)$.

We will define the locus associated to an ideal sheaf ${\cal I}_L$ via the following configuration matrix in the same ambient space.
\begin{eqnarray} \label{Ldef}
L= \left[ \begin{array}{c|cccc} \mathbb{P}^1 &1&1&1&1 \\\mathbb{P}^1 &1&1&1&1 \\\mathbb{P}^1 &1&1&1&1 \\\mathbb{P}^1 &1&1&1&1 \\\mathbb{P}^1 &1&1&1&1  \end{array} \right]
\end{eqnarray}
That is, $L$ is a one dimensional locus determined by the vanishing of four degree $(1,1,1,1,1)$ polynomials. We then have that the following sheaf restricts to $V_X$ on the Calabi-Yau threefold.
\begin{eqnarray}
V={\cal O}(2,-2,0,0,0)\otimes {\cal I}_L \oplus {\cal O}(-1,1,0,0,0)^{\oplus 2}
\end{eqnarray}

Using the same techniques which were employed in the proceeding example, we can then write down a Koszul sequence for this case,
\begin{eqnarray} \label{2kos}
0 \to \wedge^2 {\cal N}^{\vee} \otimes V \to {\cal N}^{\vee}\otimes V \to V \to V_X \to 0
\end{eqnarray}
and compute the dimensions of the cohomologies which appear. This uses the direct analogue of the sequences (\ref{idealdef1}) and (\ref{idealdef2}) for this case. In fact, in this example it turns out to simplify the computation if one also uses the Koszul sequence associated to (\ref{Ldef}).
\begin{eqnarray}\label{Lkos}
0 \to \wedge^4 {\cal N}_L^{\vee}\to \wedge^3 {\cal N}_L^{\vee}\to \wedge^2 {\cal N}_L^{\vee}\to {\cal N}_L^{\vee} \to {\cal O} \to {\cal O}_L \to 0
\end{eqnarray}
Here ${\cal N}_L={\cal O}(1,1,1,1,1)^{\oplus 4}$. This is used to compute the cohomology of ${\cal O}_L$, the trivial bundle restricted to $L$, in a straightforward fashion.

One finds the following cohomological dimensions for the sheaves appearing in (\ref{2kos})
\begin{eqnarray}
h^*(\wedge^2 {\cal N}^{\vee} \otimes {\cal O}(2,-2,0,0,0) \otimes {\cal I}_L)&=&(0,0,360,0,3,0)\\\nonumber
h^*({\cal N}^{\vee} \otimes {\cal O}(2,-2,0,0,0) \otimes {\cal I}_L)&=&(0,0,480,0,0,0)\\\nonumber
h^*({\cal O}(2,-2,0,0,0) \otimes {\cal I}_L)&=&(0,0,117,0,0,0) \; .
\end{eqnarray}
Using this data it is trivial to show that all families are of type 2 with respect to the resolution (\ref{2kos}). In addition, we have that $\wedge^3 V={\cal I}_L$ and it is simple to see from the defining sequence of ${\cal I}_L$ that $h^3(\wedge^3 V) = h^3({\cal I}_L)=0$. Therefore, in this case \vref{v-theorem3} gives us no information about any possible vanishing couplings.

\vspace{0.1cm}

Let us now compare this (negative result) to that which we obtain from considering the following "lift" of $V_X$ to the ambient space 
\begin{eqnarray}
V_{\textnormal{bundle}}={\cal O}(2,-2,0,0,0)\oplus {\cal O}(-1,1,0,0,0)^{\oplus 2} \; .
\end{eqnarray}
The bundle $V_{\textnormal{bundle}}$ also restricts to (\ref{VX2}) on the Calabi-Yau threefold. In this case, a straightforward analysis of the Koszul sequence reveals that all of the families are of type 1. We have that $h^3(\wedge^3 V_{\textnormal{bundle}})=0$ so this lift of $V_X$ shows, using \vref{v-theorem4}, that all of the Yukawa couplings between families vanish. One can, once again, confirm that this result is indeed correct in this case using the $U(1)$ symmetries associated with the line bundle model.

\vspace{0.1cm}

This example illustrates an important fact in applying the types of vanishing theorems we are discussing in this paper to the Koszul sequence. Not only will different ambient space embeddings of the Calabi-Yau threefold lead to different constraints in such a context. Considering different sheaves on the ambient space that restrict to $V_X$ on the Calabi-Yau will also lead to different textures of vanishing couplings. Especially in higher codimensions, the array of different ideal sheaves alone that you could utilize in the fashion we have employed here becomes vast, and they are presumably only a subset of the possibilities. Thus we see that we end up with a plethora of possible sources of vanishing Yukawa couplings from Koszul sequences in such contexts, even before the possible use of other resolutions are considered.

\section{Comparisons to Other Sources of Vanishing Yukawa Couplings} \label{comparisonsec}

%%%%%%%%%%%%%%%%%%%%%%%%%%%%%%%%

The results of the prior sections hopefully make clear that the explicit form of the bundle and Calabi-Yau manifold (as encoded in a resolution such as that  in \eref{seq2}) chosen in heterotic compactification can provide powerful vanishing criteria on the perturbative superpotential. It is natural to wonder how the effects described here overlap (if at all) with similar ``topological vanishing criteria'' that have been seen elsewhere in the heterotic literature? In this section we will explore vanishing conditions on Yukawa couplings due to two such phenomena 
\begin{itemize}
\item Elliptic fibration structure in the Calabi-Yau threefold (see e.g. \cite{Bouchard:2006dn,Braun:2006me}).
\item Stability structure of the holomorphic vector bundle (i.e. vanishing couplings due to ``stability walls" \cite{Kuriyama:2008pv,Anderson:2010ty,Buchbinder:2014sya}).
\end{itemize}
Note that both of these effects can be traced back to the presence of a symmetry, higher dimensional diffeomorphism invariance and gauge invariance respectively. Our goal will be to consider several examples in which the effects above are present and the resulting vanishing conditions can be compared to the results of our theorems given in previous sections.

\subsection{Vanishing tri-linear couplings on elliptically fibered manifolds}\label{leray_sec}

Recent work has highlighted the fact that the vast majority of known Calabi-Yau threefolds exhibit genus one fibrations \cite{Gray:2014fla,Anderson:2016cdu,Anderson:2016ler,Anderson:2017aux,Anderson:2018kwv,Taylor:2012dr,Johnson:2014xpa,Huang:2018gpl,Huang:2018esr,Huang:2019pne}. That is, they can be written as a fibration (i.e. a surjective morphism, $\pi: X_3 \to B_2$, for some complex surface, $B_2$) in which the generic fiber, $\pi^{-1}(b)$ for some point $b \in B_2$, is a genus one curve. Moreover, in \cite{Rohsiepe:2005qg,Gray:2014fla,Anderson:2016cdu,Anderson:2017aux} it was demonstrated that many Calabi-Yau manifolds can be written as fibrations in a multitude of distinct ways.

Just as it was observed in previous sections that structure associated to the ambient space of a complete intersection Calabi-Yau manifold can strongly constrain the heterotic couplings (as in the vanishing criteria of Sections \ref{arg1} and \ref{arg2}), so too can the ``decomposition" of a Calabi-Yau manifold into a fiber/base. This effect was first illustrated explicitly in \cite{Braun:2006me}.

As in prior sections, the origin of the constraint on couplings arises from the constrained form that cohomology must take -- this time in a fibered geometry. To make this clear, we must decompose bundle-valued cohomology groups in terms of the fiber/base of the Calabi-Yau geometry. Technically, this is achieved via a Leray spectral sequence.

Let $\pi: X_3 \to B_2$ be a genus-one fibered Calabi-Yau threefold as above and $V_X$ be any holomorphic vector bundle defined over $X_3$. Then the Leray spectral sequence\footnote{Note that in general the Leray spectral sequence can contain multiple non-trivial terms. However, in the present case, the complex dimension of the fiber is $1$, and hence $R^{m}\pi_{*} (V)$ is non-vanishing only for $m=0, 1$. This causes the spectral sequence to terminate at the first level and we will thus omit details of the spectral sequence itself in this section, only stating the result in \eref{base_decomp}.} provides a simple tool to relate the cohomology of $V_X$ over $X_3$ to some associated cohomology groups on the base, $B_2$ as
\beq\label{base_decomp}
H^p(X,V_X)=\bigoplus_{p=l+m} H^l(B_2, R^m \pi_*(V_X)) \;.
\eeq
Here $R^{m}\pi_{*}(V_X)$ is the $m$-th direct image sheaf of the bundle $V_X$  (pushed forward under the fibration $\pi$).  $R^{m}\pi_{*} (V_X)$ can be locally defined via the pre-sheaf over any open set $\mathcal{U}$ on $B$ 
\beq\label{push_forwards}
\mathcal{U} \rightarrow H^m(\pi^{-1}(\mathcal{U}), V_X|_{\pi^{-1}(\mathcal{U})}) \ .
\eeq

The form above makes it clear that the tri-linear couplings of the theory will be constrained. As in previous sections, in order for the map 
\beq
H^1(X,V_X) \times H^1(X,V_X) \times H^1(X,V_X) \to H^3(X,\wedge^3V_X)=H^3(X,{\cal O}_X)
\eeq
to give a non-zero result it must be the case that the representation of the cohomology can be re-written as a consistent/unique morphism in whatever geometry it is described in. In the present discussion, the form of \eref{base_decomp} makes it clear that the trilinear pairing must map into a top-form \emph{on the complex base surface $B_2$}. It can be verified that under push-forward to the base
\beq
H^3(X,{\cal O}_X)=H^2(B_2, R^1\pi_* {\cal O}_X)=H^2(B_2, K_{B_2})=\mathbb{C} \; .
\eeq
Thus, in order to produce a non-trivial mapping under the Yoneda pairing, it is clear that the tri-linear coupling must take the form
\beq\label{leray_form}
H^0(B_2,R^1\pi_*(V_X)) \times H^1 (B_2, R^0\pi_*(V_X))\times H^1 (B_2, R^0\pi_*(V_X)) \to \mathbb{C} \; .
\eeq
The requirement that the couplings must be ``filtered" in this way on an elliptically fibered Calabi-Yau manifold was first observed in \cite{Braun:2006me} where it was referred to as a \emph{``(p,q) Leray selection rule}". 

In the next subsection we will consider an explicit example of a bundle on a fibered Calabi-Yau manifold and the ensuing vanishing conditions on its couplings.
 
 \subsubsection{An elliptic CICY example}
 In this section our goal will be to compare the vanishing criteria induced from \eref{leray_form} to that explored in the previous sections. To that end, it is helpful to consider an example of a Calabi-Yau manifold that is \emph{both} genus one fibered and described as a complete intersection variety in a product of projective spaces. In particular, we will consider a Calabi-Yau threefold which is elliptically fibered (i.e. a genus one fibration that also admits a section $\sigma: B_2 \to X_3$) and described as
 \begin{eqnarray} \label{eg1}
X_3 = \left[ \begin{array}{c|ccc} 
\mathbb{P}^1 & 1 & 1  \\ 
\mathbb{P}^2 & 1 & 2   \\ 
\mathbb{P}^1 & 1 & 1  \\
\mathbb{P}^1 & 1 & 1 \\
\end{array}\right]\;~.
\end{eqnarray}
For this threefold $h^{1,1}(X_3)=4$ and $h^{2,1}(X_3)=50$. This manifold is elliptically fibered $\pi: X_3 \to \mathbb{P}^1 \times \mathbb{P}^1$ with the elliptic fiber described as 
 \begin{eqnarray} \label{fiber}
\left[ \begin{array}{c|ccc} 
\mathbb{P}^1 & 1 & 1  \\ 
\mathbb{P}^2 & 1 & 2   \\ 
\end{array}\right]\;~
\end{eqnarray}
(i.e. the elliptic curve defined via the anti-canonical hypersurface in $dP_1$). In addition, $X_3$ can be shown to have a non-trivial (rank one) Mordell-Weil group (see e.g. \cite{silverman}). That is, it admits two rational sections in the classes
\begin{align}
&S_1= -D_1 +D_2+D_3+D_4 \\
&S_2=2D_1-D_2 + 4D_3+4D_4
\end{align}
where $D_i$ are the restrictions of the hyperplanes in the ambient projective space factors (note that these divisor classes are related to the actual section morphisms $\sigma_i: B_2 \to X_3$ via the Shioda map \cite{Grimm:2015wda}).

Over this threefold we can define the following gauge bundle, chosen for simplicity to be a sum of line bundles
\beq\label{eg_v}
V_X=L_1 \oplus L_2 \oplus L_3={\cal O}_X(-1,1,-2,2)\oplus {\cal O}_X(-1,1,2,-2)\oplus {\cal O}_X(2,-2,0,0)  \; .
\eeq
Written in a basis defined by $S_1, S_2$ and $\pi^*(D_3), \pi^*(D_4)$ (the latter being the pullbacks of the basis of divisors in the base, $\mathbb{P}^1 \times \mathbb{P}^1$) we have
\beq
V_X={\cal O}_X(S_1)\otimes {\cal O}_X(-3D_3 +D_4)\oplus {\cal O}_X(S_1)\otimes {\cal O}_X(D_3-3D_4) \oplus {\cal O}_X(2S_1)\otimes {\cal O}_X(2D_3 + 2D_4) \; .
\eeq
This decomposition will make it particularly straightforward to calculate the higher derived push-forwards $R^i\pi_*(V_X)$ and hence the coupling constraints induced by
\eref{leray_form}.

The bundle $V_X$ in \eref{eg_v} has structure group embedded into $SU(3)$ and leads to a 4-dimensional heterotic theory with gauge group $E_6 \times U(1) \times U(1)$. A straightforward calculation yields that $h^1(X,V_X)=8$ and $h^1(X, V_X^\vee)=6$ and hence we obtain a simple, two-generation $E_6$ toy model. The bundle is chosen to be slope poly-stable within the K\"ahler cone of $X_3$ and to satisfy the anomaly cancellation conditions of the heterotic theory. 

To begin, let us consider the tri-linear couplings of the form ${\bf 27}^3$ defined as the triple product of elements of $H^1(X,V_X)$. Here, $H^1(X,V_X)=H^1(X,L_1) \oplus H^1(X,L_2)$. A straightforward calculation (see e.g. \cite{Braun:2005zv,Anderson:2015yzz} for techniques) yields that 
\begin{align}
& R^0\pi_* L_1=R^0\pi_* ({\cal O}_X(S_1)\otimes \pi^*({\cal O}_X(-3D_3 +D_4)))=R^0\pi_* ({\cal O}_X(S_1))\otimes {\cal O}(-3D_3 +D_4) \\
&  R^0\pi_* L_1=0 \\
& R^0\pi_* L_2=R^0\pi_* ({\cal O}_X(S_1)\otimes \pi^*({\cal O}_X(D_3 -3D_4)))=R^0\pi_* ({\cal O}_X(S_1))\otimes {\cal O}(D_3 -3D_4) \\
&  R^0\pi_* L_2=0 \; .
\end{align}
In the present example, it can be verified that $R^0\pi_*({\cal O}_X(S_1))={\cal O}_{B_2}$  and  $R^1\pi_* L_1=R^1\pi_* L_2=0$. Hence
\begin{align}
&H^1(X,L_1)=H^1(\mathbb{P}^1 \times \mathbb{P}^1, {\cal O}(-3D_3 + D_4)) \\
&H^1(X,L_2)=H^1(\mathbb{P}^1 \times \mathbb{P}^1, {\cal O}(D_3 -3D_4)) \; .
\end{align}
From this result, it is clear that no coupling of the form \eref{leray_form} is possible. Thus, due to the fiber/base decomposition of the Calabi-Yau threefold, we have determined that all 64 possible ${\bf 27}^3$ couplings must vanish in this example.

Interestingly, in this particular example, it is the case that the elliptic fibration criteria above actually imposes \emph{more vanishing} conditions on the Yukawa couplings than arise from the results of Sections \ref{arg1} and \ref{arg2} (i.e. constraints determined from the Koszul sequence associated to the ambient space of the CICY threefold in \eref{eg1}). For example, the line bundle $L_1={\cal O}(-1,1,-2,2)$ has cohomology $H^1(X,L_1)$ that descends from $H^2({\cal A}, {\cal O}_{\cal A}(-2,0,-3,1))$ and thus is of type $\tau=2$, as is the cohomology associated to $H^1(X,L_2)$. Since the ambient space is 5-dimensional, this coupling would in fact \emph{not} be forced to vanish by \vref{v-theorem1}. It should be noted however, that neither of these two approaches carries intrinsically more information and for another example \vref{v-theorem1} could impose more vanishing criteria than that obtained from fibration structure. The over-arching message should simply be that the geometric origins of topological vanishings criteria are myriad and all of them should be taken into account to gain a complete picture of the heterotic effective theory.

It should be observed that it is also possible to determine that these couplings must vanish due to the presence of Green-Schwarz massive $U(1)$ symmetries in the effective theory that arise from heterotic stability walls. This effect will be examined in detail in the next subsection.

Finally, it should be noted that not only do recent results indicate that the majority of known Calabi-Yau threefolds are elliptically fibered, but further that a given manifold can be described as a fibration in many ways (in \cite{Anderson:2017aux} manifolds were found with order 10s, 100s and in one case an \emph{infinite} number of inequivalent genus one fibrations). In such cases, the fibrations will impose vanishing conditions on the trilinear couplings such as \eref{leray_form} \emph{for each distinct fibration}. In general this will be a large number of constraints that imply that the majority of couplings will vanish in such models. This is very analogous to the observations of Sections \ref{arg2} and \ref{secegs} that one bundle could have multiple resolutions and thus inherit constraints on its couplings from many distinct sources. Thus, in the same spirit, the ubiquity of genus one fibrations in known datasets of Calabi-Yau threefolds further indicates the non-genericity of couplings in 4-dimensional heterotic compactifications over such manifolds.

\subsection{Vanishing tri-linear couplings due to stability walls}

In this section we will consider another effect that can constrain the form of the perturbative superpotential in heterotic theories -- namely the constraints that so-called ``stability walls" \cite{Sharpe:1998zu,Anderson:2009sw,Anderson:2009nt,Anderson:2010tc,Anderson:2010mh} place on the tri-linear couplings. These vanishing conditions have been studied extensively in \cite{Kuriyama:2008pv,Anderson:2010tc,Buchbinder:2014sya} and we will briefly summarize some of those results here and relate them to the vanishing conditions studied in this work.

Simply stated, the notion of ``stability walls" in heterotic theories is related to the conditions for $N=1$ supersymmetry in the low-energy effective theory -- namely that the vector bundle be slope poly-stable. Equivalently, due to theorems of Donaldson, Uhlenbeck and Yau \cite{Uhlenbeck,Donaldson2}, the gauge connection must satisfy the Hermitian-Yang Mills (HYM) equations
\beq
g^{a{\bar b}}F_{a{\bar b}}=0~~~~\text{and}~~~F_{ab}=0 \; .
\eeq
Here $g^{a{\bar b}}$ is the Ricci-flat metric on the Calabi-Yau manifold, $F$ is the field strength, and $a,b=1,2,3$. It is clear from the form of these equations that they manifestly depend on the K\"ahler and complex structure of the base Calabi-Yau threefold and that a solution to this equation is not guaranteed for all values of the Calabi-Yau moduli. The moduli that violate the HYM equations are ``lifted" by a combination of D- and F-term effects in the effective $N=1$ theory \cite{Anderson:2009nt,Anderson:2009sw,Anderson:2010mh,Anderson:2010tc,Anderson:2010ty} (or more generally should simply never be counted in the massless modes of the low energy theory). In particular, the constraint of stability can form an allowable sub-cone within the K\"ahler cone of the Calabi-Yau threefold. On the boundaries of this sub-cone -- called \emph{stability walls} -- the bundle moves from being stable to poly-stable and decomposes as a direct sum
\beq
V_{wall}=\bigoplus_i {\cal V}_i \label{bundle_wall}
\eeq
with ${\cal V}_i$ stable for all $i$ and satisfying $\mu({\cal V}_i)=0$. 

In the limit shown in \eref{bundle_wall} the structure group, $H$ of the vector bundle becomes reduced, leading to its commutant, $G$ within $E_8$ to be \emph{enhanced}. Such enhancements in gauge symmetry at special loci in moduli space is of course a common effect in string compactifications. In the present case the enhancement of symmetry appears as additional Abelian factors to the gauge group. For example, suppose that the initial bundle $V_X$ has structure group $H=SU(3)$ with commutant $E_6$ in $E_8$. At the special locus it could for example, split into a sum of two sub-bundles with ranks $2$ and $1$ respectively or become a sum of three line bundles. In the former case the gauge group is enhanced to $E_6 \times U(1)$ and in the latter $E_6 \times U(1) \times U(1)$.

It is these enhanced Abelian gauge symmetries that constrain the form of the $E_6$ effective theory, even though the $U(1)$ gauge fields are generically Green-Schwarz massive. As a result, it is possible to explicitly see the constraints that these enhanced Abelian gauge factors place on the effective theory -- in particular the Yukawa couplings are constrained to be gauge invariant under these enhanced $U(1)$ symmetries near the stability wall \cite{Kuriyama:2008pv,Anderson:2010tc}. Importantly, as was argued in \cite{Anderson:2010tc}, since the superpotential is a \emph{holomorphic} function of the fields, any vanishing couplings that are detected near the enhanced symmetry locus must remain zero as the theory is moved to a more general point in moduli space. Since vector bundles do in general exhibit a rich sub-cone structure due to stability (with many  ``walls" arising in higher-dimensional K\"ahler moduli spaces), this effect can contribute powerful vanishing conditions on the Yukawa couplings of heterotic theories. We turn now to a simple example of this effect and compare it to the vanishing constraints obtained in Sections \ref{arg1} and \ref{arg2}.

\subsubsection{An example monad bundle exhibiting stability walls}
As an illustration of these techniques, we consider the following $SU(3)$ bundle defined via the monad construction and first studied in \cite{Anderson:2010tc}:
\beq
0 \to V_X \to {\cal O}_X(1,1)\oplus{\cal O}_X(2,0)\oplus{\cal O}_X(3,-1)\oplus{\cal O}_X(-2,1) \stackrel{f}{\longrightarrow} {\cal O}_X(4,1) \to 0\label{mon_stab_eg}
\eeq
This bundle is defined over the Calabi-Yau hypersurface
\begin{eqnarray} \label{eg_monad_wall}
X_3 = \left[ \begin{array}{c|c} 
\mathbb{P}^1 & 2   \\ 
\mathbb{P}^3 & 4   \\ 
\end{array}\right]\;
\end{eqnarray}
with Hodge numbers $(h^{1,1},h^{2,1})=(2,86)$. Importantly for comparison with the results of this paper, this bundle is actually ambiently defined\footnote{This is clear from the fact that the four polynomial entries of the map $f$ given in \eref{mon_stab_eg} cannot all vanish simultaneously over ${\cal A}=\mathbb{P}^1 \times \mathbb{P}^3$.}, that is it is defined by restriction of the analogous monad sequence given by \eref{mon_stab_eg} defined over the ambient space $\mathbb{P}^1 \times \mathbb{P}^3$. We'll use this fact later in this section in applying the theorems from Sections \ref{arg1} and \ref{arg2} to constrain the tri-linear couplings.

To begin though, we review the constraints on the couplings obtained in \cite{Anderson:2010tc} from a stability wall analysis. As argued in Section 5 of that reference, this bundle is only stable within a sub-cone of the K\"ahler moduli space and inherits constraints from \emph{two} distinct stability walls which place different constraints on the couplings. We will summarize here the constraints from only one of these (the lower wall of \cite{Anderson:2010tc}) here where the bundle decomposes as
\beq
V_X={\cal K}_1\oplus {\cal F}_1
\eeq
where 
\beq
0 \to {\cal F}_1 \to {\cal O}(1,1)\oplus{\cal O}(2,0)\oplus {\cal O}(-2,1) \to {\cal O}(4,1) \to 0
\eeq
and ${\cal K}_1={\cal O}(3,-1)$. Near this locus in moduli space, the effective theory enhances to yield an $E_6 \times U(1)$ symmetry. Analyzing the spectrum in terms of this decomposition of the bundle then, we find the more detailed spectrum given in Table \ref{table_mon1} where subscripts denote charges under the enhanced $U(1)$ symmetry.

The first conclusion to be drawn from Table \ref{table_mon1} is that all of the ${\bf 27}$ fields have the \emph{same} $U(1)$ charge and hence all ${\bf 27}^3$ couplings vanish perturbatively.  Moreover, in \cite{Anderson:2010tc} this result was combined with the constraints from the other stability wall (and an analysis was done to determine how the textures persisted into general points in field space) which showed that only 9 out of a possible 165 ${\bf \overline{27}}^3$ couplings were possibly non-vanishing.

We can now compare these vanishing conditions to those obtained from the theorems derived in the present paper. To begin, as noted above, the bundle $V_X$ in \eref{mon_stab_eg} is ambiently defined on ${\cal A}=\mathbb{P}^1\times \mathbb{P}^3$. A standard calculation yields that
\begin{eqnarray}\label{type_decomp_mon}
h^*(X,V_X)= (0,13,9,0) \\
h^*({\cal A},V) = (0,13,0,0,0)\\
h^*({\cal A}, {\cal N}^{\vee} \otimes V) = (0,0,0,9,0) \\
h^*({\cal A},V^\vee)=(0,9,0,0,0) \\
h^*({\cal A},{\cal N}^{\vee} \otimes V^{\vee})= (0,0,0,13,0)
\end{eqnarray}
Thus, all elements of $H^1(X,V_X)$ (i.e. the ${\bf 27}$s) are of type 1. Likewise, the elements of $H^1(X, V_X^\vee)$ (i.e. the ${\bf \overline{27}}$s) are also all of type 1.  Thus, the results of \vref{v-theorem1} and \vref{v-theorem3} (applied to $V_X$ and its dual) guarantee that in this example, all ${\bf 27}^3$ and ${\bf \overline{27}}^3$ couplings must be zero! In summary then, we find for this example that these diverse sources of vanishing conditions are all consistent and agree for a subset of the possible Yukawa couplings. Moreover, the vanishing conditions arising from the theorems of Sections \ref{arg1} and \ref{arg2} are in fact \emph{stronger} in this case than those arising from a stability wall analysis and not directly tied to any obvious residual gauge symmetries of the theory.

\begin{table}[t]
\begin{center}
\begin{tabular}{|c|c|c|c|}
  \hline
  Representation & Field Name & Cohomology & Multiplicity \\ \hline
  ${\bf 27}_{-1}$ & $f_1$ & $H^1(X,{\cal F}_1)$ & 13 \Tstrut\Bstrut\\ \hline
  ${\bf \overline{27}}_{1}$ & $\bar{f}_3$ & $H^1(X,{\cal F}_1^{\vee})$ & 1 \Tstrut\Bstrut \\ \hline
  ${\bf \overline{27}}_{-2}$ & $\bar{f}_4$ & $H^1(X,{\cal K}_1^{\vee})$ & 8  \Tstrut\Bstrut\\ \hline
  $({\bf 1},{\bf 2})_{-3}$ & $C_1$ & $H^1(X,{\cal F}_1 \otimes {\cal K}_{1}^{\vee})$ & 70   \Tstrut\Bstrut\\ \hline
  \end{tabular}
\mycaption{Field content of the $E_6 \times U(1)$ theory determined by the bundle $V_X$ in \eref{mon_stab_eg} which decomposes as $V_X={\cal K}_1 \oplus {\cal F}_1$ on a stability wall.}
\label{table_mon1}
\end{center}
\end{table}

%%%%%%%%%%%%%%%%%%%%%%%%%%%%%%%

\section{Conclusions and Future Directions} \label{concsec}

So-called `topological vanishings' of Yukawa couplings in heterotic compactifications have frequently been observed in the literature. In this work, we have taken a systematic look at the geometric effects previously shown to generate such vanishings and we have added several new theorems to their ranks in Sections \ref{arg1} and \ref{arg2}.

Our primary results are the following:
\begin{itemize}
\item Vanishing Theorems \ref{v-theorem1}-\ref{v-theorem4}: These provide generalizations of previous topological vanishing theorems that have appeared in \cite{Blesneag:2015pvz,Blesneag:2016yag}. The various theorems vary in what they assume and claim. They cover a range of cases from where $V_X$ descends from a vector bundle on an ambient space, through where $V_X$ descends from an ambient sheaf, all of the way to situations that don't necessarily involve the Koszul sequence that describes such structure at all. In the most general formulation, constraints on the cubic couplings are provided based on the description of the underlying gauge bundle via some explicit resolution 
\begin{eqnarray} \label{conseq}
\ldots \to {\cal F}_2 \to {\cal F}_1 \to {\cal F}_0 \to V_X \to 0
\end{eqnarray}
in terms of sheaves ${\cal F}_i$. 
\item A series of examples in Section \ref{secegs}: These are primarily used to illustrate that a single gauge bundle $V_X$ on a given Calabi-Yau $X$ can be associated to many different resolutions of the form (\ref{conseq}). In the examples we provide these arise from embedding $X$ in different ambient spaces and also from isolating a variety of different sheaves on a single ambient space, all of which descend to $V_X$. Each of these descriptions of $V_X$ can lead to different textures of vanishing Yukawa couplings utilizing the theorems of the proceeding bullet point. Thus we observe that such vanishing theorems will lead to a plethora of complementary constraints for a given compactification.
\item A comparison to other sources of vanishing couplings in Section \ref{comparisonsec}:  Here we discuss some of the other sources of textures of vanishing Yukawa couplings that have arisen in the heterotic literature, namely those due to elliptic fibration structure and stability walls. We demonstrate that the constraints from such effects are complementary to those from the vanishing theorems mentioned above, giving distinct information about the same heterotic compactification. This leads to situations where one has an even larger set of overlapping Yukawa vanishing conditions restricting a given heterotic compactification.
\end{itemize}

\noindent In view of these results, we can consider their implications for the physics of $4$-dimensional heterotic compactifications. The first, clear consequence of the previous sections is the following observation: \\

\noindent \emph{We expect the perturbative superpotential trilinear couplings of most $4$-dimensional, ${\cal N}=1$ heterotic theories, obtained from compactifications on Calabi-Yau threefolds, to be highly constrained -- i.e. non generic}. \\

\noindent These constraints seem to arise ubiquitously \cite{Blesneag:2015pvz,Blesneag:2016yag,Gray:2019tzn}. Almost all known Calabi-Yau threefolds are elliptically fibered, and indeed often fibered in 10s or even hundreds of different ways (see e.g. \cite{Anderson:2017aux,Huang:2019pne}). Each of these different fibrations will contribute its own vanishing conditions. Picking a given Calabi-Yau at random from the known set, one finds that it is easy to  describe it as an embedding in a plethora of different ambient spaces. Gauge bundles on the threefold can, as described in Section \ref{secegs}, be represented as the restriction of a large variety of sheaves on each of those ambients. All of these different choices lead to more potential constraints on the Yukawa couplings. Most vector bundles also exhibit stability walls, which may be numerous in higher dimensional K\"ahler cones. Each of these will also lead to constraints, and the effects we have just discussed do not even include constraints from resolutions of the form (\ref{conseq}) that are not associated to the Koszul sequence.

Obviously, these textures of vanishing Yukawa couplings could play a fundamentally important role in heterotic model building, both causing and solving problems. To give one example of this, the physical quark masses in the Standard Model require a comparatively heavy top quark. Such a mass hierarchy could be achieved by topological vanishings that left only one family with \emph{perturbatively} non-zero couplings, leaving the remainder to be generated by non-perturbative effects. 

\vspace{0.2cm}

Given any set of superpotential Yukawa couplings, it is clear that some interactions could be set to zero by an appropriate change of basis on field space.
One key question to ask given the results presented in this paper is whether the vanishing of Yukawa couplings that have been presented could simply be viewed as finding a suitable basis change in this fashion. The answer to this is {\it definitely not} in general. We will illustrate this here for just one of the many vanishing conditions described in this work: the case of a bundle defined over an elliptically fibered space, $\pi: X_3 \to B_2$ which pushes forward to a bundle on the base, $B_2$. As described in Section \ref{comparisonsec}, in this case, the bundle valued harmonic 1-forms, elements of $H^1(X, V_X)$, are `filtered' according to their behavior under pushforward. For an $SU(3)$ bundle, the only possible non-vanishing ${\bf 27}^3$ coupling must be of the form shown in \eref{leray_form}:
\beq\label{leray_form2}
H^0(B_2,R^1\pi_*(V_X)) \times H^1 (B_2, R^0\pi_*(V_X))\times H^1 (B_2, R^0\pi_*(V_X)) \to \mathbb{C}
\eeq

Let us label elements of $H^1(X,V_X)$ that are associated to elements of $H^1 (B_2, R^0\pi_*(V_X))$, as $x_i$ where $i=1,\ldots, n_x$ and elements associated to $H^0(B_2,R^1\pi_*(V_X))$ as $y_a$ where $a=1,\ldots, n_y$. The total number of matter fields will be denoted $n=n_x+n_y$. From the form of the product given in \eref{leray_form2}, it is clear that no coupling can exist except for those of the form $xxy$ and its permutations. We can thus ask: how many vanishing couplings does this correspond to? There are a total of $\frac{1}{3!}n(n+1)(n+2)$ independent Yukawa couplings. Since all except $n_x n_y$ of these vanish we have a total number of independent vanishing couplings given by
\begin{eqnarray}
\frac{1}{6}n(1+n)(2+n)-n_x(n-n_x)\;.
\end{eqnarray}
This quantity is minimized at fixed $n$ for $n_x=n/2$, thus this value gives the minimum number of vanishing couplings that will be present in any model. Using this value for $n_x$ we obtain the following number of constraints:
\begin{eqnarray} \label{mindude}
\frac{1}{12}n(4+3n +2 n^2)
\end{eqnarray}
Note that $n/2$ may not be an integer and so the above minimum number of vanishing independent Yukawa couplings can be viewed as a conservative lower bound.

It only remains to compare this number to the largest number of independent Yukawa couplings that can be arranged to vanish by a basis change on field space. Obviously this cannot be larger than $n^2$ - the number of degrees of freedom in a basis change matrix itself. In fact, given that the overall scale in the matrix cannot affect vanishings, $n^2-1$ independent zeros can be obtained maximally in this fashion. So all that remains is to compare this value with the minimum number of enforced vanishings (\ref{mindude}) from the condition \eref{leray_form2}. Doing a straightforward comparison we arrive at the following, conservative, conclusion: {\it A heterotic $E_6$ model defined on an elliptically fibered Calabi-Yau geometry, $\pi: X_3 \to B_2$ for which $V_{X_3}$ pushes forward to a bundle on $B_2$ has vanishing superpotential Yukawa couplings that cannot be explained by a change of basis of the fields if the number of ${\bf 27}$ matter fields is $n>3$.}

\vspace{0.2cm}

There are a number of natural extensions of this work that it would be interesting to investigate in the future. The extension to bundles of higher rank should pose no practical problems and has simply been omitted here to try and prevent the discussion from becoming overly complex. Another interesting extension of this work arises in the interpretation of the perturbative, trilinear superpotential couplings in heterotic theories as objects in the deformation theory of the underlying bundle. For instance, as studied in \cite{Berglund:1995yu,Anderson:2011ty}, the cubic singlet couplings ${\bf 1}^3$ can be viewed as higher order obstructions in the bundle moduli space (whose first order space of deformations is counted by $h^1(X,\textnormal{End}_0(V_X))$). In that case, the deformations in question \emph{preserve} the rank of the bundle. In the present case of $SU(3)$ bundles, the trilinear couplings between \emph{charged} matter fields are linked to deformations/obstructions that \emph{change the rank} of the gauge bundle, i.e. to deformations of an augmented bundle
\beq
V_X^{SU(3)} \oplus {\cal O}_X^{\oplus m}
\eeq
where $m=0,\ldots 5$. In other words, viewing the chosen $SU(3)$ bundle as a special point in the moduli space of an $E_8$ bundle, the charged matter of the theory can parametrize rank-changing deformations which move in this broader moduli space. These deformations can in turn be obstructed at various higher orders. In this context, it would be interesting to understand if the vanishing conditions we are studying here correspond to some underlying `protected directions' in the moduli space of $E_8$ bundles, analogous to the Tian-Todorov theorem \cite{tian,todorov} which protects the infinitesimal deformation space of Calabi-Yau 3-folds from higher order obstructions. While it is known that the first order deformations of $SU(n)$ bundles on Calabi-Yau \emph{can} be fully obstructed \cite{Thomas:1999tq}, the structure of higher order obstructions to the $E_8$ moduli space in this context remains relatively unexplored. 

Another natural question arising from the present work is what consequences do these observations have in the context of string dualities? For example, recent work on the ubiquity of elliptic fibrations in Calabi-Yau 3-folds and 4-folds has made clear that the space of string effective theories which can be realized by \emph{both} heterotic and F-theory compactifications is potentially much larger than anticipated. In this context then, once we consider heterotic vanishing conditions like the one described in detail above for elliptically fibered manifolds, it is natural to ask what the consequences of this result are for F-theory? Naively, F-theory calculations of Yukawa couplings do not seem to offer an obvious source of such topological vanishings and instead seem generally generic in form (see e.g. \cite{Cvetic:2019sgs} for a recent perspective). However, many such investigations have thus far been undertaken without a full global understanding of G-flux in 4-dimensional compactifications of F-theory and its contributions to the superpotential. In the future it would be valuable to explore these vanishing conditions in some explicit heterotic/F-theory dual pairs and to determine what G-flux must be present and if these vanishings extend away from the stable degeneration limit.

Finally, an obvious question in the context of Yukawa couplings in heterotic theories is how many of these vanishings will persist to higher order in the theory or when including contributions from non-perturbative effects. While non-perturbative effects do not have a simple integral form as shown here, at least some of these geometric features (for example, constraints arising from stability walls as reviewed in Section \ref{comparisonsec}) may have consequences for such couplings as well. Recent progress in the development of tools for determining these non-perturbative contributions (e.g. \cite{Buchbinder:2019eal}) may make the exploration of such questions more tractable in the future.

\section*{Acknowledgements}

The work of L.A and J.G. are supported in part by NSF grant PHY-2014086. M.L., M.M. and R.S. are funded in part by Vetenskapsr\aa det,  under grant numbers 2016-03873, 2016-03503, and 2020-03230.

%%%%%%%%%%%%%%%%%%%%%%%%%%%%%%%%%%%%%%%%%%%%%%%%%%%%

\end{document}